\documentclass[conference]{IEEEtran}

\newif\ifAnon\Anonfalse

\usepackage[hyphens]{url}
\PassOptionsToPackage{bookmarks,hidelinks}{hyperref}
\usepackage{hyperref}
\hypersetup{breaklinks=true}
\usepackage{tugraz_defaults}

\usepackage{csquotes}
\usepackage{longtable}

\usepackage{graphicx}
\usepackage{caption} 
\usepackage{subcaption}
\usepackage{pgfplotstable}
\usepackage{hyperref}
\usepackage[skins]{tcolorbox}

\renewcommand{\paragraph}[1]{\vspace{0.0cm}\noindent\textbf{#1}\ }

\usepgfplotslibrary{dateplot}
\usepgfplotslibrary{fillbetween}

\makeatletter
    \newenvironment{myalign*}{%
      \setlength{\abovedisplayskip}{-0.2\baselineskip}%
      \setlength{\abovedisplayshortskip}{\abovedisplayskip}%
      \start@align\@ne\st@rredtrue\m@ne
    }%
    {\endalign}
\makeatother

\tcbset{skin=enhanced,
    arc=0pt,
    left=3pt,
    right=3pt,
    fonttitle=\bfseries,
    colback=white,
    interior style={white,opacity=0.9},
    frame style={blue,solid},
    leftrule=1pt,
    rightrule=1pt,
    bottomrule=1pt,
    segmentation style={black,solid,opacity=0.2,line width=1pt}
}

\begin{document}

\date{}

\title{\Large \bf Remote Memory-Deduplication Attacks}

\ifAnon
\author{}
\else
\author{
{\rm Martin Schwarzl}\\
Graz University of Technology
\and
{\rm Erik Kraft}\\
Graz University of Technology
\and
{\rm Moritz Lipp}\\
Graz University of Technology
\and
{\rm Daniel Gruss}\\
Graz University of Technology
}
\fi

\ifAnon
    \author{}
\else
    \author{\IEEEauthorblockN{Martin Schwarzl}
    \IEEEauthorblockA{Graz University of Technology\\
    martin.schwarzl@iaik.tugraz.at}
    \and
    \IEEEauthorblockN{Erik Kraft}
    \IEEEauthorblockA{Graz University of Technology\\
    erik.kraft5@gmx.at}
    \and
    \IEEEauthorblockN{Moritz Lipp}
    \IEEEauthorblockA{Graz University of Technology\\
    moritz.lipp@iaik.tugraz.at}
    \and
    \IEEEauthorblockN{Daniel Gruss}
    \IEEEauthorblockA{Graz University of Technology\\
    daniel.gruss@iaik.tugraz.at}}
\fi

% \IEEEoverridecommandlockouts
% \makeatletter\def\@IEEEpubidpullup{6.5\baselineskip}\makeatother
% \IEEEpubid{\parbox{\columnwidth}{
%     Network and Distributed Systems Security (NDSS) Symposium 2022\\
%     27 February – 3 March 2022, San Diego, CA, USA\\
%     ISBN 1-891562-66-5\\
%     https://dx.doi.org/10.14722/ndss.2022.23081\\
%     www.ndss-symposium.org
% }
% \hspace{\columnsep}\makebox[\columnwidth]{}}

\maketitle

\begin{abstract}
Memory utilization can be reduced by merging identical memory blocks into copy-on-write mappings.
Previous work showed that this so-called \emph{memory deduplication} can be exploited in local attacks to break ASLR, spy on other programs, and determine the presence of data, \ie website images.
All these attacks exploit memory deduplication across security domains, which in turn was disabled.
However, within a security domain or on an isolated system with no untrusted local access, memory deduplication is still not considered a security risk and was recently re-enabled on Windows by default.

In this paper, we present the first fully remote memory-deduplication attacks.
Unlike previous attacks, our attacks require no local code execution. %
Consequently, we can disclose memory contents from a remote server merely by sending and timing HTTP/1 and HTTP/2 network requests.
We demonstrate our attacks on deduplication both on Windows and Linux and attack widely used server software such as Memcached and InnoDB.
Our side channel leaks up to \SI{34.41}{\byte/\hour} over the internet, making it faster than comparable remote memory-disclosure channels. %
We showcase our remote memory-deduplication attack in three case studies:
First, we show that an attacker can disclose the presence of data in memory on a server running Memcached.
We show that this information disclosure channel can also be used for fingerprinting and detect the correct libc version over the internet in \SI{166.51}{\second}.
Second, in combination with InnoDB, we present an information disclosure attack to leak MariaDB database records.
Third, we demonstrate a fully remote KASLR break in less than \SIx{4} minutes allowing to derandomize the kernel image of a virtual machine over the Internet, \ie 14 network hops away.
We conclude that memory deduplication must also be considered a security risk if only applied within a single security domain.
\end{abstract}

\section{Introduction}
Memory deduplication is a widely used technique to reduce memory utilization by detecting physical pages with the same content and merging them.
Merged pages are marked as read-only and copy-on-write.
If one of the merged pages is modified, a copy-on-write page fault is triggered, and the page is again copied to a new physical location.
With the introduction of Windows 8.1, memory deduplication had become a default feature~\cite{Yosifovich2017}.
On Linux, kernel-same-page merging is used by kernel-virtual machines or if the \texttt{madvise} syscall is used with a flag indicating that the page is mergeable.

Previous work demonstrated memory-deduplication attacks performed by a local attacker in both local environments (\ie local native code execution) and the cloud (\ie local code execution in a virtual machine)~\cite{Suzaki2011,Barresi2015} exploiting page combining on Windows and kernel-same-page merging on Linux.
Memory-deduplication attacks can detect co-location in the cloud~\cite{Suzaki2011}, hide communication in virtualized environments~\cite{Xiao2012covert,Xiao2013security}, fingerprint operating systems~\cite{Owens2011}, fingerprint websites via JavaScript~\cite{Gruss2015dedup} and break ASLR on Linux as well as on Windows by exploiting pages with almost fixed content~\cite{Barresi2015}.
Bosman~\etal\cite{Bosman2016} leveraged memory deduplication in combination with Rowhammer to escape from a browser sandbox.
Razavi~\etal\cite{Razavi2016} used memory deduplication to facilitate Rowhammer attacks on co-located virtual machines.
Palfinger~\etal\cite{Palfinger2019Prying} demonstrated that memory deduplication can also be exploited in file systems like ZFS.
Lindemann~\etal\cite{Lindemann2018A} demonstrated efficient fingerprinting via memory deduplication in co-located virtual machines.
In concurrent work, Kim~\etal\cite{Kim2021Breaking} showed a KASLR break on virtual machines on VMWare ESXi.
Following the recommendation of all these attack papers, memory deduplication was disabled on Linux and Windows by default.

More recently, vendors switched to more fine-grained security policies.
Windows 10, for instance, again enables page combining by default but restricts it to only deduplicate within a security domain but not across security domains, stopping existing attacks.
We also observe that the popular Ubuntu 20.04 Linux distribution enables kernel-same-page merging by default for KVM-based virtual machines.
Memory-deduplication attacks with local code execution are considered out of scope in their threat model.
Systems without local code execution (native or in a virtual machine) for the attacker can still be considered secure with these mitigation strategies.
However, it remains unclear whether remote attacks \emph{without local code execution} are possible.

Our work faces three challenges which have to be solved to perform remote memory-deduplication attacks:

\begin{compactitem}
    \item \textit{C1: Remotely amplify latencies for non-repeatable events.}
Remote timing attacks require high latencies in the side channel to deal with noisy networks.
Page-fault-type interrupts cannot be arbitrarily repeated (e.g., for copy-on-write page faults, the page is copied and writeable after the page fault).
Hence, existing amplification techniques are not directly applicable.

All previous memory deduplication attacks focused on cross-domain deduplication. 
Deduplication within one domain is considered secure (Windows re-enabled it for that reason). 
Intra-domain deduplication is visible outside of the domain if the timing latency is exposed over a web server or public API to the attacker domain.

\item \textit{C2: Trigger and observe copy-on-write pagefaults in a victim domain that shares no memory with any attacker domain.}
All previous memory deduplication attacks require local code execution (in native or sandboxed code).
Remote requests are usually not held in memory for a long time. 
To speed up, the access of frequent data, in-memory caching mechanisms like Memcached are used in websites.

\item \textit{C3: Find remote request paths that do not only keep attacker-controlled data in memory but also provide the attacker with control over alignment and in-memory representation.}
To enable byte-by-byte leakage, a target is required that allows alignment changes as described by Bosman~\etal\cite{Bosman2016}.
\end{compactitem}

In this paper, we solve the mentioned challenges and demonstrate the \textbf{first fully remote memory-deduplication attacks}, just using requests to an HTTP web server.
Our attacks infer timing differences caused by copy-on-write page faults on the server from the latency of network requests and responses.
We demonstrate attacks on default-configured and fully updated Windows (native) and Linux (virtual machines) installations using default-configured standard server software such as Memcached.
We measure the capacity of our side channel in a remote covert channel scenario and achieve a transmission rate of \SI{302.16}{\byte/\hour} in a local area network and \SI{34.41}{\byte/\hour} over the internet, which is faster than comparable remote memory-disclosure channels (\eg NetSpectre~\cite{Schwarz2019netspectre} achieved \SI{7.5}{\byte/\hour} in a local area network).

We demonstrate three different remote memory-deduplication attacks, illustrating the potential of our technique.
In the first attack, we disclose the presence of data on a remote server running Memcached.
The information disclosure works by uploading data blobs into the key-value store, freeing the deduplicated item, getting the same item reassigned, and triggering a copy-on-write page fault by modifying the page's content.
We also exploit this information disclosure channel for fingerprinting, \ie which shared libraries are used on the remote system.
Our attack detects the correct libc version over the internet in \SI{166.51}{\second}.

In the second attack, we present a fully remote KASLR break on a virtual machine running on a remote cloud machine.
By targeting kernel pages that contain kernel addresses but have all remaining bytes of the page fixed, we can successfully derandomize the kernel offset of a Linux virtual machine.
We show that we can not only mount this attack in a local area network setting using HTTP/1 but, moreover, leverage HTTP/2 to successfully break KASLR on a server that is \SIx{14} network hops away within \SIx{4} minutes.
We emphasize that vendor responses to local KASLR breaks are often that KASLR is only meant as a mitigation for remote attacks.

In a third attack, we disclose database records byte-by-byte from a MariaDB database server with an InnoDB storage engine.
Our attack works by crafting requests that create byte misalignments within target pages, allowing byte-wise content guessing.
This attack is particularly dangerous as it leaks attacker-unknown memory contents from a remote server, similar as in powerful Spectre attacks~\cite{Kocher2019,Schwarz2019netspectre}.
We can leak \SI{1.5}{\byte/\hour} in a local area network.

We conclude that memory deduplication must also be considered a security flaw if only applied within a security domain and even if local attackers are excluded from the threat model.
As our attacks are full remote attacks, we emphasize that the remote attack vector has to be mitigated as well.
Consequently, we responsibly disclosed all of our attacks to the corresponding vendors and work with them on finding mitigations before the public release of this paper.
We will open-source our tools on GitHub with the conclusion of the responsible disclosure~\footnote{\url{https://github.com/IAIK/Remote-Page-Deduplication-Attacks}}.

\subheading{Responsible Disclosure.}
We responsibly disclosed our findings to Microsoft, Red Hat, Canonical, and AWS, on February 8th, 2021.
The issues are tracked under CVE-2021-3714.%

\subheading{Contributions.} The main contributions of this work are:
\begin{compactenum}
\item We present the first fully remote memory-deduplication attacks and show that these must be considered a security flaw even if only applied within a security domain.
\item We show that we can remotely fingerprint shared libraries to infer the exact versions via Memcached in-memory databases.
\item We present a fully remote KASLR break on a Linux virtual machine running in the cloud within only \SIx{4} minutes.
\item We demonstrate a fully remote byte-by-byte memory disclosure attack on a MariaDB database server with an InnoDB storage engine, leaking \SI{1.5}{\byte/\hour}.
\end{compactenum}

\subheading{Outline.}
The remainder of the paper is organized as follows.
In~\cref{sec:background}, we provide the required background about memory deduplication and remote timing attacks.
In~\cref{sec:threat_model}, we state a threat model and provide an attack overview.
In~\cref{sec:attack_primitives}, we present the attack primitives that we use for remote memory-deduplication attacks.
In~\cref{sec:case-studies}, we evaluate the performance of our remote memory-deduplication attacks in three case studies on Windows and Linux, targeting Memcached, MariaDB (with InnoDB), and the Linux kernel.
In~\cref{sec:mitigations}, we discuss the results and state-of-the-art mitigations for remote memory-deduplication attacks.
We conclude in \cref{sec:conclusion}.

\section{Background}\label{sec:background}

In this section, we provide background on memory deduplication, memory-deduplication attacks, and remote timing attacks, as well as Address Space Layout Randomization.

\subsection{Memory Deduplication}\label{sec:bg_pagededuplication}
Sharing memory is not only crucial for inter-process communication but also to reduce memory utilization and cache pressure.
Modern operating systems use different techniques to use shared memory whenever possible.
For instance, when creating a new process with \texttt{fork()}, the memory is marked as \emph{copy-on-write}, meaning that it is first shared between parent and child process and only copied (\ie duplicated) when one of the processes attempts to write to it.
This is implemented by marking the memory read-only and raising a page fault upon a write access.
Another example is the loading of any type of file (including, \eg a program or library binary files).
The operating system keeps files in the page cache and maps them into all processes that request access.

Neither of these approaches leads to the deduplication of identical but dynamically generated memory pages.
Hence, operating systems have introduced content-based memory deduplication, which regularly scans the entire physical memory for pages with identical content.
All but one of the identical pages are released, while the remaining one is marked as copy-on-write.
Content-based memory deduplication has traditionally been applied across all security domains on all major operating systems.
On Windows, the mechanism is called page combining~\cite{Yosifovich2017} and kernel same-page merging on Linux~\cite{Arcangeli09Increasing}.
However, security research has revealed that this enables a range of attacks, as we discuss in the next sub-section.

\subsection{Memory-Deduplication Attacks}\label{sec:bg_pagededuplication_attacks}
In a memory-deduplication attack, the attacker first generates candidate pages for deduplication.
If the attacker guesses the content of a page in memory fully correctly, it is deduplicated.
Until the deduplication took place, the attacker repeatedly writes to the candidate pages (without changing the content).
As soon as the deduplication took place, this triggers a copy-on-write page fault, increasing the access latency drastically.
Hence, the access latency reveals whether a victim process had a page with the exact same content, \ie memory deduplication forms a content-probing oracle.

The first memory-deduplication attack, demonstrated by Suzaki~\etal\cite{Suzaki2011}, was used to detect applications running in other virtual machines.
Owens~\etal\cite{Owens2011} also exploited memory deduplication to fingerprint the operating system version via unique pages per operating system in virtual machines.
Gruss~\etal\cite{Gruss2015dedup} showed that memory-deduplication attacks are possible from JavaScript running on a website opened in a browser.
Barresi~\etal\cite{Barresi2015} demonstrated that it is possible to break address space layout randomization (ASLR) on both Windows and Linux using memory deduplication.
Razavi~\etal\cite{Razavi2016} exploited memory deduplication to perform Rowhammer attacks on applications in virtualized environments.
Bosman~\etal\cite{Bosman2016} used memory-deduplication attacks to create more sophisticated exploits and used the ASLR break via memory deduplication to create an end-to-end JavaScript exploit which leverages Rowhammer to achieve arbitrary memory read and write. 
Oliverio~\etal\cite{Oliverio2017Secure} proposed a mitigation against active memory-deduplication attack called VUsion, which enforces same behavior when accessing shared and non-shared pages, a write-xor-fetch policy, and random memory allocation.
Lindemann~\etal\cite{Lindemann2018A} showed another fingerprinting attack to detect co-location in virtual machines.

Palfinger~\etal\cite{Palfinger2019Prying} showed that memory deduplication can also be leveraged in file systems like ZFS to fingerprint the operating system in the cloud of co-located machines.
In concurrent work, Kim~\etal\cite{Kim2021Breaking} demonstrated a KASLR break on VMWare ESXi.

\subsection{Remote Timing Attacks}\label{sec:remote_attacks}
Timing attacks were heavily researched in the last two decades.
Since network connections are getting more and more stable, at higher transmission rates, as well as lower and more consistent latencies, remote timing attacks have become increasingly interesting for attack research.
Brumley and Boney~\etal\cite{Brumley2003Remote} demonstrated that it is possible to extract SSL private keys over a local area network.
Ac\i{}i\c{c}mez~\etal\cite{Aciicmez2007d} attacked AES via a remote cache based attack.
In 2009, Crosby~\etal\cite{Crosby2009} showed the possibilities of remote timing attacks and how to reliably determine the number of requests required to distinguish certain timing differences over the network.
There were several remote timing attacks on AES~\cite{Zhao2009cache,Jayasinghe2010remote,Aly2013attacking,Saraswat2014} following Bernstein's idea of attacking AES~\cite{Bernstein2005}.
Van Goethem~\etal\cite{VanGoethem2015} exploited timing side channels in browsers. 
Irazoqui~\etal\cite{Irazoqui2015Lucky} showed that it is possible to exploit cache timing differences in TLS in a local area network.
Van Hoef~\etal\cite{VanHoef2016Heist} leveraged TCP windows to observe the exact size of a cross-origin resource.
Van Goethem~\etal\cite{VanGoethem2020Timeless} showed that remote timing attacks can be performed over the Internet by exploiting concurrency in HTTP/2 and observing the order the packets return, which depends on the server-side timing, instead of the client-side timing.
Kurt~\etal\cite{Kurt2020Netcat} showed that Data Direct I/O can be used in combination with Remote Direct Memory Access to spy on keystrokes during SSH sessions.

More closely related to our work is Schwarz~\etal\cite{Schwarz2019netspectre}, who showed that Spectre attacks are possible over the network if certain gadgets exist on the target system.
Similar to the most powerful attacks we present, they can leak arbitrary data from an execution context.
They achieve a leakage rate of up to \SI{7.5}{\byte/\hour}, which can be sufficient to leak a cryptographic key over the time frame of multiple hours.

\subsection{Address Space Layout Randomization}\label{sec:background:kaslr}
To exploit memory corruption bugs, the knowledge of addresses of specific data is often required since address randomization is applied in both user space and kernel space.
Over the past years, different side-channel attacks allowed to reduce the entropy of the randomization or to break it entirely.
Hund~\etal\cite{Hund2013} measured the execution time of page-fault handling to observe which kernel addresses are mapped and thus cached in the TLB.
Jang~\etal\cite{Jang2016} used hardware transactional memory to observe the same effect.
Other software-based side channel attacks exploited predictors~\cite{Evtyushkin2016ASLR,Lipp2020takeaway}, side channels introduced by mitigations against other attacks~\cite{Canella2020kaslr}, the power consumption of the processor~\cite{Lipp2021Platypus}, and other microarchitectural properties~\cite{Gruss2016Prefetch, Koschel2020, Goktas2020blind}, even from JavaScript~\cite{Gras2017aslr, Canella2019Fallout}.
As a consequence of these local attacks on KASLR, operating system vendors but also parts of the academic community considered KASLR only as a defense against remote attackers.
In remote attacks, KASLR indeed is still considered a valuable line of defense since the attacker cannot as easily probe the address space as with local attacks.

\section{Threat Model \& Attack Overview}\label{sec:threat_model}
In our threat model, the attacker has no ability to execute code on the target machine: not natively, not in a virtualized environment, and also not via JavaScript~\cite{Bosman2016,Gruss2015dedup} or another scripting language.
However, the attacker can provide attacker-controlled content to the remote target, \eg a network request the attacker sends to the host with content the attacker controls.

We assume that the victim keeps the attacker-controlled content in RAM.
This occurs, for instance, if the attacker sends network requests that are cached in request pools, or binary large objects provided to a web application and later on stored in a database or cached in a buffer.

We assume that memory deduplication techniques are active on the victim's machine.
We emphasize that this is the case under default settings on current Ubuntu Linux installations (kernel-same-page merging for virtual machines) and on current Windows installations (page combining).

We make no assumptions about software bugs, for instance, memory safety violations in the applications we analyze.

\begin{figure}[t]
    \centering
    \resizebox{\hsize}{!}{
        \tikzsetnextfilename{attack-overview}
        \input{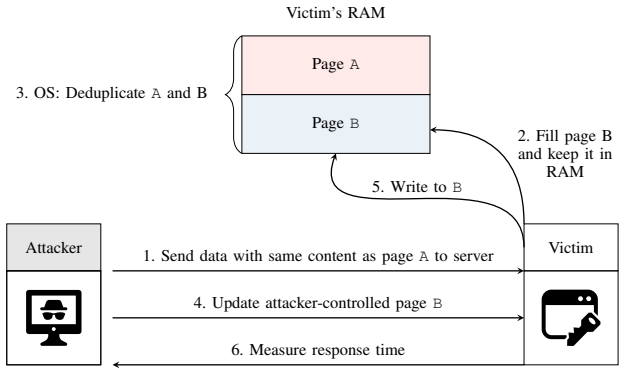}
    }
    \caption{Overview of a remote memory-deduplication attack.}
    \label{fig:attack_overview}
\end{figure}

\paragraph{Attack Overview.}
Six steps are required to perform a remote memory-deduplication attack as illustrated in~\cref{fig:attack_overview}.
First, the attacker sends a request to the victim with a page of data (page \texttt{B}) containing the same content as a page already present in memory (page \texttt{A}).
Afterwards, the attacker waits for some time until the two pages are merged by the operating system and point to the same physical address.
Next, the attacker updates the attacker-controlled data and triggers a page-fault on the victim application.
Depending on the response time of the victim, the attacker observes whether the page was deduplicated or not.

\paragraph{Difference to already presented attacks.}
All of the previous presented attacks~\cite{Suzaki2011,Owens2011,Gruss2015dedup,Barresi2015,Bosman2016} require local code execution via a native binary or JavaScript and co-location to the victim's machine.
Remote memory-deduplication attacks extend the scope by enabling attacks on remote web servers by exploiting an API that allows uploading of attacker-controlled data and place it into the main memory such that it might be deduplicated.
Comparison of state-of-the-art memory deduplication attacks to our work is listed in~\cref{tab:requirements}.
While some of the techniques shown by previous work are similar, we solved those challenges for memory deduplication attacks in the context of a remote attacker. 
As evidenced by other fully remote attacks~\cite{Schwarz2019netspectre,VanGoethem2020Timeless}, specific timing requirements and the applicability to many-hop internet connections, remain a challenge that is only solved for specific cases. 

\begin{table*}[t]
\setlength{\aboverulesep}{0pt}
\setlength{\belowrulesep}{0pt}
    \begin{center}
      \adjustbox{max width=\hsize}{
        \begin{tabular}{p{0.12\hsize}|p{0.09\hsize}|p{0.15\hsize}|p{0.03\hsize}|p{0.10\hsize}|p{0.16\hsize}|p{0.14\hsize}}
        \toprule
        \textbf{Attacks}    & \textbf{Location}                      & \textbf{Environment}                 & \textbf{Local} & \textbf{Type} & \textbf{Attack Type} & \textbf{Performance}\\
        \midrule
        Suzaki~\etal\cite{Suzaki2011} & Co-located                   & Cross-VM (Cloud) & Yes & Native binary & 
        Fingerprinting & -
        \\
        Owens~\etal\cite{Owens2011}   & Co-located                  & Cross-VM (Cloud) & Yes & 
        Native binary & Fingerprinting & -
        \\
        Gruss~\etal\cite{Gruss2015dedup}  & Remote               & Browser/Cross-VM (Cloud) & Yes & JavaScript & Fingerprinting & -
        \\
        Barresi~\etal\cite{Barresi2015}  & Remote              & Cross-VM (Cloud) & Yes & Native binary & ASLR break    & \SI{8.7}{days}
        \\
        Bosman~\etal\cite{Bosman2016}   & Remote               & Browser (Same-machine) & Yes & JavaScript & Bytewise leakage, ASLR break, Rowhammer  & \SI{2.75}{\hour}
        \\
        Lindemann~\etal\cite{Lindemann2018A}   & Co-located               & Cross-VM (Cloud) & Yes & Native binary & Fingerprinting  & \SI{1.8}{\hour}
        \\ 
        Kim~\etal\cite{Kim2021Breaking} & Co-located               & Cross-VM (Cloud) & Yes & Native binary & KASLR break  & \SI{12}{\minute}
        \\
        \textbf{Our work}    & Remote         & \textbf{Internet/Local-NW}  & \textbf{No} & \textbf{None} &
        Bytewise leakage, KASLR break, Fingerprinting & \SI{1.5}{\byte/\hour} (Local-NW) / \SI{4}{\minute} / \SI{166.51}{\second}
        \\
        \bottomrule \\
      \end{tabular}
    }
    \\
        {\footnotesize \qquad Location: Attacker's location \qquad Local: local code execution \qquad Type: Type of local code execution \qquad Perf: Reported Attack Performance }

  \end{center}
  
  \caption{Comparison of state-of-the-art memory deduplication attacks. }
  \label{tab:requirements}

\end{table*}

\section{Attack Primitives}\label{sec:attack_primitives}
In this section, we describe our basic attack primitives and define the requirements for a remote attacker to perform a fully remote memory-deduplication attack without execution of any attacker-controlled code on the victim system.

The main primitives for our attack are memory deduplication being enabled, a web service/API that lets a remote attacker read/modify data stored in RAM and an accurate remote timer that allows distinguishing the round-trip time of the network packets.

\subsection{Memory Deduplication}\label{sec:attack_primitives:memory_deduplication}
\paragraph{Page combining.}
Page combining was introduced in Windows 8.1. 
On Windows, a special kernel thread scans over the whole memory to detect pages that have identical content~\cite{Yosifovich2017}.
This scan is triggered about every \SIx{15} minutes on Windows 10~\cite{Yosifovich2017}.
If pages with identical content are found, the pages are combined to a single page to save memory.
The page-table entries of the pages then point to one of the two pages, which is then shared across processes and marked as read-only and copy-on-write.
When writing to this shared page, a copy-on-write fault occurs, and a new copy of the page is created for the writing process~\cite{Yosifovich2017}.

Page combining can easily be disabled via the Windows registry or using Powershell, \eg using the \texttt{Disable-MMAgent} command.
Page combining was temporarily disabled for security reasons after several memory-deduplication attacks were discovered~\cite{Barresi2015,Bosman2016,Gruss2015dedup}.
However, page combining was re-enabled on Windows more recently and is active on desktop machines by default, as well as on server machines if the \texttt{full} terminal server role is enabled.
In addition, a Windows 10 process has the possibility to disable page combining~\cite{WindowsProcessMitigationSideChannel}.

We observed this effect by checking all terminal server options in Windows 2016 (Version 1607, Build 14393.693) and Windows Server 2019 (Version 1809, Build 17763.737).
We also empirically validated that for Windows 10 Professional 20H2 19042.746 and Windows 10 Home 19041.746 page combing was active by default.
On Microsoft's Azure Cloud~\cite{Microsoft2019Azure} it is also possible to acquire such Windows Server VMs with this configuration.
We created a Windows 2019 Server (Version 1809, Build 17763.1697) and can also confirm that page combining is enabled after setting the full terminal server role.
On Windows, it is also possible to force page combining using the \texttt{RtlAdjustPrivilege} and \texttt{NtSetSystemInformation} functions. %

\paragraph{Linux Kernel Same-Page Merging}
Kernel-Same-Page Merging (KSM) is the counterpart of page combining on Linux~\cite{Arcangeli09Increasing,RedHatKSM}.
KSM is enabled and mainly used for Kernel Virtual Machine (KVM) virtualized machines, for instance, on Red Hat Linux~\cite{RedHatKSM}.
On Ubuntu 20.04, we observed that when \texttt{qemu-system-common} with KVM support is installed on a host machine, \texttt{KSM\_ENABLED} is set to \texttt{AUTO} in \texttt{/etc/default/qemu-kvm}, enabling KSM per default for non-virtualized instances.
We also set up an Ubuntu 20.04 server image and observed the same behavior after installing QEMU.
Like on Windows, a kernel thread scans over the memory and merges pages with identical content to a single page, which is then marked as copy-on-write~\cite{RedHatKSM}.

\begin{figure}[t]
    \centering
    \resizebox{\hsize}{!}{
        \tikzsetnextfilename{num-of-pages-dedup-time}
        \begin{tikzpicture}
            \begin{axis}[
            enlarge x limits={0.01},
            style={font=\footnotesize},
            width={\hsize},
            height=3cm,
            xlabel={pages\_to\_scan},
            ylabel={Seconds},
            xmin=0,
            ymin=0,
            xmode=log,
            log ticks with fixed point,
            ]
            \addplot[scatter=true,mark=o,nodes near coords*={\Label},visualization depends on={value \thisrow{label} \as \Label},]  table[x index = {0}, y index = {1}, col sep=comma]{data/pages_to_scan_timing_diff/results.txt};
            \end{axis}
            
\end{tikzpicture}
    }
    \caption{The deduplication time of a single \SI{4}{\kilo\byte}-page strongly depends on the number of pages\_to\_scan (sleep\_millisecs=200). }
    \label{fig:dedup_time}
\end{figure}
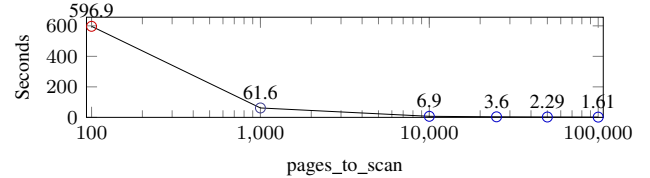

On Linux, only pages are merged that are marked as mergeable, \ie using the \texttt{madvise} syscall and setting the \texttt{MADV\_MERGEABLE} flag~\cite{Arcangeli09Increasing}.
This is the default for pages of KVM virtual machines.
The user can configure how many pages should be scanned per invocation (\texttt{pages\_to\_scan}).
The default value on a Ubuntu 20.04 is $100$ \texttt{pages\_to\_scan} in a time interval of \SI{200}{\milli\second}.
Therefore, in the optimal case, up to \SIx{500} \SI{4}{\kilo\byte} pages can be deduplicated per second.
\Cref{fig:dedup_time} illustrates the required time for a single page being deduplicated, with a different number of pages\_to\_scan set, and the default value of \SI{200}{\milli\second} for sleep\_millisec.
We evaluate the deduplication time for a single page depending on the scanned pages on a remote server equipped with an Intel Xeon E3-1240 running Ubuntu 20.04.
However, it is recommended to increase the number of \texttt{pages\_to\_scan} to increase the deduplication performance~\cite{SUSE2021KSM}.
The tool KSMtuned sets the time interval per default to \SI{10}{\milli\second} and increases \texttt{pages\_to\_scan} to \SIx{1250}~\cite{RedHatKSM}.
This would lead to a maximum \SI{512}{\mega\byte} being deduplicated per second.
We asked a \textbf{cloud provider}, which hosts multiple hundred thousand websites, for the KSM config used in \textbf{production}.
The cloud provider uses a configuration of \texttt{sleep\_millisecs=30,pages\_to\_scan=500}, leading to at maximum \SI {65.84}{\mega\byte} (16500 pages) being deduplicated.
The average time after a single page is deduplicated with that configuration is \SI{34.57}{\second} ($n=10,\sigma=6.3\%$).

\subsection{Service/Web API.}
We assume that the victim machine provides network-accessible services, \eg a REST API, enabling users to store and modify data blobs.
There are no restrictions in the way these data blobs are controlled, \ie the user could either upload and replace files or send strings to the server, as long as the memory location of the data blob does not change.

\subsection{Remote Timer.}
To get the best possible low-latency timing information, we use the hardware timestamps from the network interface card.
We measure the timing difference between the last packet sent and the first response byte received from the server (tcp\_flags=PUSH,ACK).

The victim side (which the attacker cannot control) runs under default configuration.
However, on the attacker side (that is under full control of the attacker), we disable the following optimizations in the Linux network parsing \texttt{sudo ethtool -K enp3s0 tso off gso off gro off}.
These options disable offloading of TCP packets to the network interface card.
Offloading might influence the timestamps on the attacker (receiver) side.
We observed for some network interface cards that due to receiver side packet coalescing, the TCP receive timestamp of the first received packet might be overwritten.
To ensure that no coalescing happens, we developed a kernel module which disables packet coalescing on the receiver side, for network cards which have this problem.

\paragraph{Network Timestamps.}
We found that one of the bottlenecks of remote attacks is the limited number of HTTP requests which can be sent using a simple HTTP requests library like \texttt{pyrequests}.
Therefore, we use asynchronous IO mechanisms to increase the number of requests per second.
Furthermore, we observed that Wireshark's TCP-field \texttt{tcp.time\_delta} reflects the timing difference between copy-on-write pages and non copy-on-write pages best.
This field calculates the timing difference between two captured packets.
Compared to the network timestamp read from the NIC, we require only \SIx{20} requests instead of \SIx{40} to distinguish $16$ overwritten copy-on-write pages over $14$ hops in the internet to build a histogram.

\subsection{Attack Setup.}\label{sec:attack_setup}
For all our case studies, we use the following setup for our local and remote scenario.

\paragraph{Local Scenario.}
The local victim machine uses an i7-6700K processor with Ubuntu 20.04 (kernel 5.4.0) and runs QEMU 4.2.1 with KVM support enabled and virtualization extensions enabled. 
Co-located in the same local area network, we have our attacker machine, which also uses an i7-6700K processor and Ubuntu 20.04 (kernel 5.4.0).
For the Linux setup, we host a virtual machine with KVM running Ubuntu Server 20.04 LTS (kernel 5.4.0-53-generic).

\paragraph{Remote Scenario.}
In addition, we used a remote Linux server by Equinix~\cite{Equinix}, running on an Intel Xeon E3-1240 CPU.
We installed the same virtual machine on the Linux server.
For our Linux machine, which was located in Amsterdam, we observed \texttt{14} network hops.

We created a virtual machine on Microsoft Azure of size Standard D4s v3~\cite{Microsoft2019Azure} and set up a Windows Server 2019 (Version 1809, Build 17763.1697) with page combining enabled.
We observed \texttt{28} hops, using the nmap traceroute command, between our network and the Windows 2019 server virtual machine, which was located in Amsterdam.
We use the same attacker machine from our local setup to perform the internet attacks.

\paragraph{Settings.}
To estimate the highest possible capacity of our remote covert channel, we try to reduce the noise as far as possible, \ie by fixing the CPU frequency of the KVM virtual machine.
To enable full scans on a moderate CPU utilization, we set the value of \texttt{/sys/kernel/mm/ksm/pages\_to\_scan} to \SIx{100000}. 
The \texttt{/sys/kernel/mm/ksm/sleep\_milliseconds} remains at the default value of \SI{200}{\milli\second}.
Furthermore, we set the CPU performance governor to performance using the \texttt{cpupower} tool to avoid noise from wake-up delays.
We later on use the default configuration of Ubuntu, Windows, and the cloud provider to calculate the leakage rates for the cases studies.

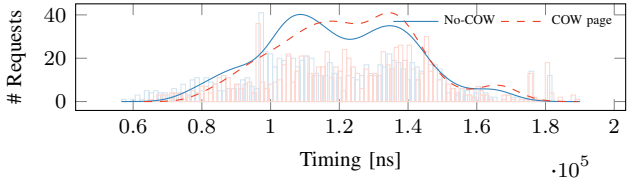
\begin{figure}[t]
  \centering
      \tikzsetnextfilename{timing-difference-single}
      \begin{tikzpicture}
\begin{axis}[
style={font=\footnotesize},
xlabel={Timing [ns]},
ylabel={\# Requests},
y label style={align=center,text width=2cm},
width=\hsize,
height=3cm,
legend style={at={(1.0,1.0)}, anchor=north east, legend columns=2, font=\tiny,draw=none,fill=none},
]

\addplot[blue!20, pattern=north east lines, fill=none, hist, hist/bins=100] table [y index=0, col sep=comma] {data/local_ampl1_1000_nocow.csv};
\addplot[red!20, fill=none, hist, hist/bins=100] table [y index=0, col sep=comma] {data/local_ampl1_1000_cow.csv};
\end{axis}

\begin{axis}[
style={font=\footnotesize},
xlabel={Timing [ns]},
ylabel={},
y label style={align=center,text width=2cm},
width=\hsize,
axis y line=none,
axis x line=none,
height=3cm,
legend style={at={(1.0,1.0)}, anchor=north east, legend columns=2, font=\tiny,draw=none,fill=none},
]

\addplot[blue, fill=none] table [x=CorrectX, y=CorrectY, col sep=comma] {data/local_nw_ampl1_distribution.csv};
\addlegendentry{No-COW};
\addplot[dashed,red, fill=none] table [x=IncorrectX, y=IncorrectY, col sep=comma] {data/local_nw_ampl1_distribution.csv};
\addlegendentry{COW page};
\end{axis}

\end{tikzpicture}
  \caption{Timing distribution of a single deduplicated page of a virtual machine in a local area network scenario on Linux KVM ($n=1000$). }
  \label{fig:timing_difference_single}
\end{figure}
\begin{figure}[t]
  \centering
      \tikzsetnextfilename{timing-difference-inet}
      \begin{tikzpicture}
\begin{axis}[
style={font=\footnotesize},
xlabel={Timing [ns]},
ylabel={\# Requests},
y label style={align=center,text width=2cm},
width=\hsize,
height=3cm,
legend style={at={(1.0,1.0)}, anchor=north east, legend columns=2, font=\tiny,draw=none,fill=none},
]

\addplot[blue!20, pattern=north east lines, fill=none, hist, hist/bins=100] table [y index=0, col sep=comma] {data/inet_ampl1_1000_nocow.csv};
\addplot[red!20, fill=none, hist, hist/bins=100] table [y index=0, col sep=comma] {data/inet_ampl1_1000_cow.csv};
\end{axis}

\begin{axis}[
style={font=\footnotesize},
xlabel={Timing [ns]},
ylabel={},
y label style={align=center,text width=2cm},
width=\hsize,
axis y line=none,
axis x line=none,
height=3cm,
legend style={at={(1.0,1.0)}, anchor=north east, legend columns=2, font=\tiny,draw=none,fill=none},
]

\addplot[blue, fill=none] table [x=CorrectX, y=CorrectY, col sep=comma] {data/inet_ampl1_distribution.csv};
\addlegendentry{No-COW};
\addplot[dashed,red, fill=none] table [x=IncorrectX, y=IncorrectY, col sep=comma] {data/inet_ampl1_distribution.csv};
\addlegendentry{COW page};
\end{axis}

\end{tikzpicture}
  \caption{Timing distribution of a single deduplicated page of a virtual machine in the internet(14 hops) ($n=1000$).}
  \label{fig:timing_difference_single_inet}
\end{figure}
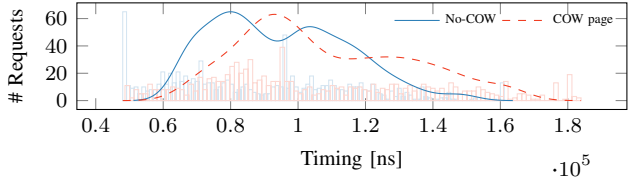

\paragraph{Evaluation.}
For a single page, we measure a local timing difference directly in the virtual machine (KVM) and observe that the average local timing difference between a regular write memory access and a memory access causing a copy-on-write page fault is \SI{7209.3}{\nano\second} ($n = 100$, $\sigma_{\textnormal{COW}}$ = $26.23\%$, $\sigma_{\textnormal{NOCOW}}$ = $29\%$) using a local timer.
We evaluate the timing difference in our local area network and on the internet using a simple HTTP server with a key-value store.
\cref{fig:timing_difference_single} illustrates the timing difference for a single page accessed with a copy-on-write page fault and a normal write access in a local area network.
In the local area network, we observe a mean timing difference of \SI{4353.91}{\nano\second} ($n=1000$).
\cref{fig:timing_difference_single_inet} shows the timing difference for a single page accessed with a copy-on-write page fault and a normal write access from our Linux server on the internet (14 hops).
While those two distributions overlap, they can be clearly distinguished in the mean respectively median values if enough samples are taken.
In addition, the timing difference can be amplified by overwriting multiple copy-on-write pages in a single request.

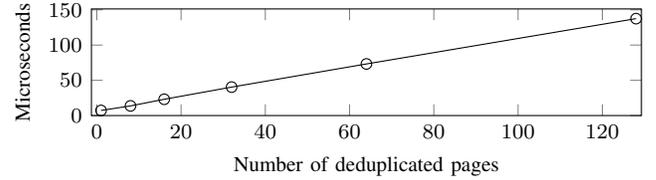
\begin{figure}[t]
  \centering
   \tikzsetnextfilename{amplification-diff}
   \begin{tikzpicture}
            \begin{axis}[
            enlarge x limits={0.01},
            style={font=\footnotesize},
            width={\hsize},
            height=3cm,
            xlabel={Number of deduplicated pages},
            ylabel={Microseconds},
            xmin=0,
            ymin=0,
            ]
            \addplot[mark=o]  table[x index = {0}, y expr =\thisrow{Y}/1000, col sep=semicolon]{data/amplification_local.csv};
            \end{axis}
            
\end{tikzpicture}
 \caption{Timing difference between amplified pages.}
 \label{fig:amplification}
 \end{figure}

\paragraph{Amplification.}
In the following paragraph we solve \textit{C1: (Remotely amplify latencies for non-repeatable events.)}.
A copy-on-write page fault can be amplified if multiple pages belonging to the same semantic entity (i.e., pages of an image file) get duplicated at the same time~\cite{Gruss2015dedup}.
Therefore, we can amplify the timing difference between multiple deduplicated pages by sending a single request, writing to those which trigger the copy-on-write, and responding back.
To evaluate the timing differences of multiple copy-on-write page faults, we evaluate a different set of pages, which triggers the page fault.
We define a test set in our local KVM machine with a test set of 1, 8, 16, 32, 64, and 128 deduplicated pages and measure the average timing difference between triggering a copy-on-write page fault and a regular write access.
We repeat the experiment \SIx{100} times and calculate the difference between the average times, which is plotted in~\cref{fig:amplification}.
We can see that there is a linear increase in terms of the timing difference with the increase of the number of deduplicated pages.
For instance, with $8$ pages, we get an average timing difference of \SI{13610.82}{\nano\second} and with $16$ pages, it is on average \SI{22946.14}{\nano\second}.

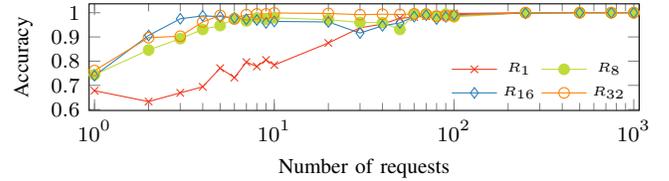
\begin{figure}[t]
  \centering
     \tikzsetnextfilename{success-over-amplification}
     \begin{tikzpicture}
            \begin{axis}[
            enlarge x limits={0.01},
            style={font=\footnotesize},
            width={\hsize},
            height=3cm,
            xlabel={Number of requests},
            ylabel={Accuracy},
            xmode=log,
            legend style={font=\tiny},
            legend style={at={(1.0,0.0)}, anchor=south east, legend columns=2, font=\tiny,draw=none,fill=none},
            grid style=dashed
            ]

            \addplot[red,mark=x]  table[x index={0},y index = {1}, col sep=semicolon]{data/box/results/meas1_kvm_ampl.csv};
            \addlegendentry{$R_{1}$}
            
            \addplot[green,mark=*]  table[x index={0},y index = {1}, col sep=semicolon]{data/box/results/meas8_kvm_ampl.csv};
            \addlegendentry{$R_{8}$}
            
            \addplot[blue,mark=diamond]  table[x index={0},y index = {1}, col sep=semicolon]{data/box/results/meas16_kvm_ampl.csv};
            \addlegendentry{$R_{16}$}
            
            \addplot[orange,mark=o]  table[x index={0},y index = {1}, col sep=semicolon]{data/box/results/meas32_kvm_ampl.csv};
            \addlegendentry{$R_{32}$}

            \end{axis}
\end{tikzpicture}
 \caption{Success rate of the classifier using the box test with a different number of deduplicated pages ($R_x$).}
 \label{fig:success_ampl}
 \end{figure}

Next, we evaluate the effect of amplification in our local area network setup with KVM.
We use the term amplification factor to indicate the number of additional pages used to amplify the signal.
We sample \SIx{1000} times and fit a CDF(cumulative distribution function) for each of the amplification factors ($1,8,16,32$) and randomly sample from the CDF.
To discover the number of requests required to achieve an accuracy higher than 95\% percent, we perform the box test by Crosby~\etal\cite{Crosby2009}.
\cref{fig:success_ampl} illustrates the number of network requests required to achieve a certain success rate for a different number of pages deduplicated by the server.
In this idealized setup, we observe that $10$ requests with an amplification factor of $8$ are enough to achieve a 95\% confidence of distinguishing write accesses on a deduplicated page (incurring a page fault) and a non-deduplicated page in a local area network if amplification is used.
The number of requests required is in a similar range for the local area network observed by Van Goethem~\etal\cite{VanGoethem2020Timeless}.

\begin{tcolorbox}[title={C1}]
\textbf{Remotely amplify latencies for non-repeatable events.}
\tcblower
  We showed the applicability of memory-deduplication attacks within the same security domain.
  We can amplify the timing differences for the copy-on-write page faults arbitrarily by leveraging the deduplication of multiple pages belonging to the same semantic entity.
  If the attacker is in control of overwriting the data, multiple copy-on-write page faults increase the latency.
\end{tcolorbox}

\paragraph{R/W bit stays cleared.}
On Windows and Linux with page combining respectively kernel-same-page merging, we observe that when a page is deduplicated, and a write access occurs to one of the corresponding virtual pages, the remaining mappings of the same physical page remain marked as \textbf{copy-on-write}. 
We empirically validate this in an experiment, where we first map two pages $A$ and $B$ with identical content and wait for deduplication.
We then write to page $A$ and thus trigger a copy-on-write page fault.
Subsequently, we analyze the R/W bit of the page-table entries for both pages and see that the R/W bit remained cleared for page $B$.
This observation is especially useful when the attacker can align data, as was shown by Bosman~\etal\cite{Bosman2016}.
In \cref{sec:case-studies:mariadb}, we exploit this behavior to amplify a single copy-on-write request via Memcached.

\section{Remote Covert Channel}\label{sec:remote_covert_channel}
For our evaluation on both Windows and Linux, we first create a covert channel using our remote memory-deduplication channel.
For this purpose, we implement a small HTTP/1.1 server in C++ to maximize the performance.
We evaluate this attack on a local-area network with a hardware switch between the attacker and the victim.

\paragraph{Capacity.}
We build a covert channel to measure the performance of our remote memory-deduplication attack in a local area network scenario.
The victim system for our transmission hosts a website that allows storing and updating files.
The website keeps the files in in-memory storage, \ie in RAM.

The sender and receiver upload an identical large file to the website hosted on the victim system.
Both use a \SI{4}{\kilo\byte} page in this large file to encode a `1'-bit.
To transmit a `1'-bit, the sender puts the same page into RAM by updating the file via the website.
The page is deduplicated with the page in the receiver's file.
Conversely, to transmit a `0'-bit, the sender modifies the page in its file such that it is not deduplicated.
The receiver sends a network request that either triggers a copy-on-write page fault or not.
With measured round-trip time, the receiver distinguishes between a `1' and a `0'.
The transmission can be parallelized in our setup by storing multiple bits at once and evaluating them in parallel.

\paragraph{Local Area Network.}
We transmit a random secret that is \SIx{8} bytes long, and repeat the experiment $100$ times.
On each repetition, we re-randomize a new \SI{8}{\byte} secret.
We observed that the Python capturing library has problems correctly parsing the packets when performing too many requests asynchronously on our webserver, we always leak 2 bytes (16 bit) in parallel for stable results.
Between the send and receive process, a delay of \SI{3}{s} was used to wait for deduplication.

In our Linux setup using amplification of $16$, we achieve an overall performance of \SI{302.16}{\byte/\hour} ($n=100,\sigma=5.81\%$), with an error rate of \SI{0.6}{\percent}. 

\paragraph{Internet.}
We run the same experiment as for the local area network.
On Linux, we used \SIx{20} requests per bit and used an amplification factor of \SIx{16} pages.
On Windows, we used \SIx{20} requests per bit and an amplification factor of \SIx{32} pages.

On the Linux server, we achieve an overall performance of \SI{34.41}{\byte/\hour} ($n=100,\sigma=5.87\%$) with an error rate of \SI{0.83}{\percent}.
On the Windows server, we use constant triggering of memory deduplication and a delay of \SI{50}{\milli\second} and achieve an overall performance of \SI{26.64}{\byte/\hour} ($n=100,\sigma=0.69\%$) with an error rate of \SI{0.18}{\percent}.
We use this number to calculate the timing for the actual wait time on Windows of \SIx{15} minutes until the deduplication succeeded, which is \SI{0.4}{\byte/\hour}.

Using the same methodology as state-of-the-art work~\cite{Barresi2015,Bosman2016}, we simulate the covert's channel performance for the default configuration of \SIx{100} pages\_to\_scan on Linux.
As it takes \SI{596.9}{\second} on the Equinix server to perform a full scan, the covert channel's performance shrinks down to \SI{0.59}{\byte/\hour}.
For the provided numbers of the cloud provider, the covert channel would achieve \SI{20.62}{\byte/\hour}.
These numbers are in a higher range as previous work, with the additional overhead of TCP, compared to the UDP sockets used in a similar attack scenario~\cite{Schwarz2019netspectre}.
The other remote timing attacks did not provide concrete numbers on their covert channel~\cite{Zhao2009cache,Jayasinghe2010remote,Aly2013attacking,Saraswat2014,Aciicmez2007d,VanGoethem2020Timeless}.

\section{Case Studies}\label{sec:case-studies}
In this section, we evaluate three case studies and demonstrate what types of attacks are possible with remote memory-deduplication.
First, we demonstrate that we can exploit remote memory-deduplication in Memcached to fingerprint the system, including the operating system.
We successfully detect the correct libc library over the internet in \SI{166.51}{\second}.
Second, we demonstrate a fully remote KASLR break by exploiting remote memory-deduplication within \SIx{4} minutes.
Third and finally, we demonstrate how to leak database records byte-by-byte from InnoDB used in MySQL and MariaDB.
In the following subsections we show how to solve C2, and C3.

\subsection{Memcached}\label{sec:case-studies:memcached}
Memcached is a fast in-memory database offering a key-value store for applications~\cite{memcached_website}.
The memory is managed using a slab allocator. 
A slab consists of a single or multiple memory pages, which are contiguous in physical memory.
Memcached always allocates a \SI{1}{\mega\byte} region and splits it into smaller chunks of equal size~\cite{memcached_website}.
Chunks or objects with a similar object size get assigned to a certain slab class.
For instance, if a slab class is \SI{64}{\byte}, the \SI{1}{\mega\byte} page is split into \SIx{16384} chunks.
Newly inserted data is assigned to the smallest slab class that the data fits in~\cite{memcached_website}.
This means a certain slab class contains objects of a certain size and assigns the objects to a chunk.
A key-value pair is managed by the \texttt{item} structure, a linked list that contains the size of the key, the value of the object, and some more metadata~\cite{memcached_website}.
Each slab class has a free list, which is a linked list~\cite{memcached_website}.
If an item gets freed, its former location is moved to the head of the free list.

\paragraph{Memory Management.}
The key-value pairs are stored contiguously in memory, which is ideal for triggering memory deduplication.
We analyzed and profiled the source code of Memcached to check which functions are used and how the memory allocation works internally.
In contrast to our expectations, Memcached does not perform an in-place replacement of the value to update.
Even with the same key used in \texttt{memcached\_set} and \texttt{memcached\_replace} operations, a new location is assigned to the updated value.
This new location is either an available free slab item from the head of the free list (do\_slabs\_alloc) or a new slab item.

After all input data from the new item is read, the old item is unlinked, and the new location linked for the item.
The old location is freed and inserted to the head of the slab's free list (code path is from complete\_nread $\to$ do\_item\_link).
If a fixed memory size is reached, a least-recently-used (LRU) eviction policy is applied on a slab-base~\cite{MemcachedLRU}
This means that ``old'' items are replaced by more frequent items in a certain slab class.

\subsubsection{Attack.}
Our basic remote memory-deduplication attack on Memcached works as follows on Linux and Windows:
First, the attacker places the targeted pages into the key-value store with a specific identifier.
Then, the attacker waits some amount of time (delay) such that the pages are deduplicated.
The deduplicated content can be for instance a static unique binary page of a specific version of the C standard library or other static binary pages in the system.
After the delay, the attacker creates a new dummy item with the same key, which puts the deduplicated target page on the free list of Memcached.
Then, the attacker updates the same item, which causes a copy-on-write page fault on the deduplicated page which is now overwritten.

\paragraph{Alignment.}
In general, it is not guaranteed that allocating memory with \texttt{malloc} internally uses \texttt{mmap} for a specific allocation size (this may depend on the libc variant, \ie glibc MMAP\_THRESHOLD is 128 kB,system configuration, and operating system versions of the victim system).
Thus, it is also not guaranteed that the allocated \SI{1}{\mega\byte} region is aligned to any specific offset.
Using \texttt{mmap} would ensure a page alignment, meaning the page offset would always be \SIx{0}.
However, in our experiments, we observed that malloc always used \texttt{mmap} internally for the \SI{1}{\mega\byte} allocations on a default configured Ubuntu Linux installation.

It is also not guaranteed that the attacker inserts the first item in the slab class, which also causes an unknown alignment as also other chunks might be inserted on the \SI{1}{\mega\byte} page.
To overcome this limitation we propose a method to generate chunks of all different sizes possible for a slab class.
We calculate all possible offsets the chunk could have on the page for a certain slab class.
These possible offsets can be computed for each possible chunk per page $i$ as \texttt{offset}:

{
\begin{myalign*}
(malloc\_offset + i \cdot chunk\_size + item\_header\_size \\ + key\_size) \mod 4096
\end{myalign*}
}%

where the \texttt{key\_size} is attacker-controlled, the \texttt{chunk\_size} depends on the slab class, $\texttt{malloc\_offset}=16$ and the size of the item header is defined as $\texttt{item\_header\_size}=56$.
Hence, to overcome the alignment issue, we use the same page with the different offsets to cover all possible alignments, which is guaranteed to include the correct alignment required for deduplication.

\paragraph{LRU.}
In a real-world application, Memcached can be expected to be heavily used by other users as well.
However, we still need to keep the data inside the data store.
We achieve this by frequently accessing the data using GET requests on the service to avoid being evicted by the LRU eviction strategy.
Note that this does not trigger copy-on-write page faults as we only read the data but do not modify it.
We discuss the eviction in more details in an attacker scenario in~\Cref{sec:appendix:memcached-eviction}.

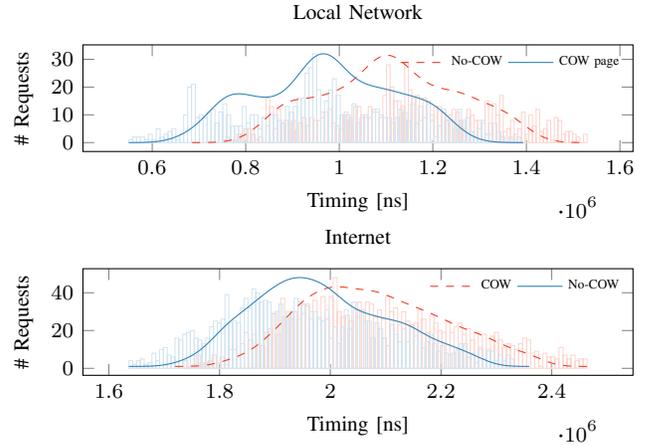
\begin{figure}[t]
  \centering
     \begin{subfigure}{\hsize}%
         \tikzsetnextfilename{memcached-p2p}
         \begin{tikzpicture}
\begin{axis}[
style={font=\footnotesize},
xlabel={Timing [ns]},
ylabel={\# Requests},
y label style={align=center,text width=2cm},
width=\hsize,
title=Local Network,
height=3cm,
legend style={at={(1.0,1.0)}, anchor=north east, legend columns=2, font=\tiny,draw=none,fill=none},
]

\addplot[red!20, fill=none, hist, hist/bins=100] table [y index=0, col sep=comma] {data/fingerprinting_local_cow.csv};
\addplot[blue!20, pattern=north east lines, fill=none, hist, hist/bins=100] table [y index=0, col sep=comma] {data/fingerprinting_local_nocow.csv};
\end{axis}

\begin{axis}[
style={font=\footnotesize},
xlabel={Timing [ns]},
ylabel={},
y label style={align=center,text width=2cm},
width=\hsize,
axis y line=none,
axis x line=none,
height=3cm,
legend style={at={(1.0,1.0)}, anchor=north east, legend columns=2, font=\tiny,draw=none,fill=none},
]

\addplot[dashed,red, fill=none] table [x=CorrectX, y=CorrectY, col sep=comma] {data/fingerprinting_local_distribution.csv};
\addlegendentry{No-COW};
\addplot[blue, fill=none] table [x=IncorrectX, y=IncorrectY, col sep=comma] {data/fingerprinting_local_distribution.csv};
\addlegendentry{COW page};
\end{axis}

\end{tikzpicture}%
     \end{subfigure}
     \begin{subfigure}{\hsize}%
         \tikzsetnextfilename{memcached-inet}
         \begin{tikzpicture}
\begin{axis}[
style={font=\footnotesize},
xlabel={Timing [ns]},
ylabel={\# Requests},
y label style={align=center,text width=2cm},
width=\hsize,
title=Internet,
height=3cm,
legend style={at={(1.0,1.0)}, anchor=north east, legend columns=2, font=\tiny,draw=none,fill=none},
]

\addplot[red!20, fill=none, hist, hist/bins=100] table [y index=0, col sep=comma] {data/fingerprinting_inet_cow.csv};
\addplot[blue!20, pattern=north east lines, fill=none, hist, hist/bins=100] table [y index=0, col sep=comma] {data/fingerprinting_inet_nocow.csv};
\end{axis}

\begin{axis}[
style={font=\footnotesize},
xlabel={Timing [ns]},
ylabel={},
y label style={align=center,text width=2cm},
width=\hsize,
axis y line=none,
axis x line=none,
height=3cm,
legend style={at={(1.0,1.0)}, anchor=north east, legend columns=2, font=\tiny,draw=none,fill=none},
]

\addplot[dashed,red, fill=none] table [x=CorrectX, y=CorrectY, col sep=comma] {data/fingerprinting_inet_distribution.csv};
\addlegendentry{COW};
\addplot[blue, fill=none] table [x=IncorrectX, y=IncorrectY, col sep=comma] {data/fingerprinting_inet_distribution.csv};
\addlegendentry{No-COW};
\end{axis}

\end{tikzpicture}%
     \end{subfigure}
 \caption{Histogram of the network requests in a local area network and in the internet setup using 16 pages to amplify the results in Memcached.}
 \label{fig:histogram_memcached}
\end{figure}

\paragraph{Evaluation.}
We evaluate our attack on Memcached 1.6.8 and connect to the Memcached service using UNIX sockets.
We evaluate this scenario using a PHP site (version 7.4.3), which is hosted on an Nginx server (version 1.18).
Our evaluation uses the local area network setup, and we also run on a Linux server on the internet 14 hops away.
Our victim server and attacker setup are the same as described in~\cref{sec:remote_covert_channel}.

We alternate between pages that do not trigger a copy-on-write page fault and pages that trigger a copy-on-write page fault due to deduplication.
In addition, we alternate the order to avoid a potential bias, which could be introduced by a fixed request order.
In total, we perform \SIx{1000} HTTP requests. 
\cref{fig:histogram_memcached} shows the timing differences we observe in this setup.
We can see that it is easy to distinguish between deduplicated pages and non-deduplicated pages.

\paragraph{Libc Fingerprinting.}
Operating system and library fingerprinting is a good starting point for penetration testing to determine potential vulnerabilities on the identified operating system or the running applications.
Those results observed from Memcached can be used to perform fingerprinting of operating systems by looking at fixed memory pages as was proposed by Owens~\etal\cite{Owens2011}.
We use the same setup as before and try to fingerprint the exact standard C lib (libc) version.
In our experiment, we probe $3$ different versions of the libc.
We perform $20$ subrequests for each version we probe on Memcached.
Our information disclosure attack detects the correct version in one sample within $44.28$ seconds ($n=100$,$\sigma=0.19\%$) and an accuracy of $90\%$, depending on which library is mapped on the victim.
Memcached can be used as an additional possibility to force deduplication and evaluate the response time, which we show in \cref{sec:case-studies:mariadb}.
The timing differences for the correct library guesses in PHP via Memcached can be seen in~\Cref{fig:php_timing_fingerprinting} (\Cref{sec:appendix-timingdiffphp}).

\paragraph{Internet.}
We run the same experiment with the same setup (Nginx, PHP, Memcached) over the internet targeting the Equinix Linux VM \texttt{14} hops away.
We detect the correct version in one sample within $166.51$ seconds ($n=100$,$\sigma=9.67\%$) and an accuracy of $90\%$. 
Using the default settings for KSM, the attack would take \SI{3.36}{\hour}.
With the settings provided by the cloud provider, the attack would take \SI{0.22}{\hour}.

\begin{tcolorbox}[title={C2}]
\textbf{Trigger and observe copy-on-write page faults in a victim domain that shares no memory with any attacker domain.}
\tcblower
With our attack on PHP-Memcached hosted on an Nginx server, we demonstrated that it is possible to trigger copy-on-write page faults within the same security domain without relying on shared memory with the attacker domain.
This can be used to perform operating system fingerprinting like was shown via Memcached over the internet.
\end{tcolorbox}

\subsection{Breaking KASLR Remotely}\label{sec:case-studies:kaslr}
By randomizing the location of kernel code, data, and drivers at every boot, KASLR makes the exploitation of memory corruption bugs in the kernel much harder (\cref{sec:background:kaslr}) as an adversary needs to guess the addresses for the attack correctly.
In the past, different side-channel attacks allowed to reduce the entropy of the randomization or to break it entirely~\cite{Hund2013,Jang2016,Evtyushkin2016ASLR,Lipp2021Platypus,Lipp2020takeaway,Canella2020kaslr,Canella2019Fallout,Gruss2016Prefetch,Koschel2020,Gras2017aslr,Goktas2020blind}.

While Klein and Pinkas~\cite{Klein2019} used an information leak in IP headers to break KASLR, so far, no remote side-channel attack has been demonstrated against KASLR.
In this section, we exploit memory deduplication to break KASLR of a virtual machine remotely.
Concurrent work by Kim~\etal\cite{Kim2021Breaking} demonstrated a KASLR break on co-located machines on VMWare ESXi break via memory deduplication within 12 minutes.

We describe the necessary building blocks and threat model to mount the attack targeting one virtual machine over the network. %

\subsubsection{Attack Scenario \& Attacker Model}\label{sec:case-studies:kaslr:attack}

We assume that the version of the operating system running on the victim machine is known to the attacker. That memory deduplication is active and enabled by the operating system (or hypervisor).
This information can be obtained by an information leak or a fingerprinting attack, similar to the one described on Memcached in~\cref{sec:case-studies:memcached}.

\subsubsection{Attack \& Building Blocks}\label{sec:case-studies:kaslr:building-blocks}

Finding the content of memory pages that are identical to the ones used by the victim operating system forms the basis of our KASLR break.
If the content of the attacker-controlled page is identical, the hypervisor deduplicates it. Thus, a subsequent write to the page yields a higher execution time forming the side-channel we exploit throughout this paper.
While a page with the same content as a kernel page allows fingerprinting the operating system, data and pointers stored on the page either change during runtime or are randomized on every boot and are, thus, less predictable.

However, on Linux, the text segment is mapped between the 1 GB region of \texttt{0xffff ffff 8000 0000} and \texttt{0xffff ffff c000 0000}.
As the kernel is 2 MB aligned, there are only 512 possible offsets in this region where the kernel can be placed.
If we find kernel pages that only contain kernel addresses and static values, \ie data that is not modified during runtime, we can generate 512 different versions of the page.
Each version corresponds to one possible offset and contains the kernel addresses if the kernel would be mapped to said offset.

A page on the victim machine is now filled with a possible content candidate.
The remote attacker uses the API provided by the victim machine to set the content of a page.
Depending on the configuration of the hypervisor on the target machine, the adversary waits until pages should be deduplicated.
Now the adversary writes to the same page using the API.
The adversary measures the time it takes to write to the page, \ie the time it takes for the network request to be handled.
If the content set by the adversary matches the targeted kernel page, the hypervisor has deduplicated the pages, and to handle the write. They have to be duplicated again.
Thus, if the content matched, the adversary observes a higher timing.
For all of the 512 different possibilities, the adversary performs these measurements, yielding a single candidate that corresponds to the currently used randomization offset.
To deal with measurement noise, the adversary has to repeat these measurements.

In addition, it is possible to amplify the side-channel leakage.
Instead of a single kernel page, multiple different kernel pages can be generated based on the assumed kernel offset and set at the same time.
Thus, instead of a single deduplication, the adversary observes multiple ones within a single measurement.

\paragraph{Finding Suitable Kernel Pages.}\label{sec:case-studies:kaslr:building-blocks:finding}
To send the content of possible kernel pages, the adversary first needs to scan possible page candidates.
This can be done upfront in an offline phase and used for kernels of the same version, thus, one assumption is that the adversary knows the version used by the victim.

To find possible page candidates, we walk the page table levels of the Linux kernel and inspect the content of each mapped \SI{4}{\kilo\byte} page.
We know in which region the text segment can be mapped and check each possible position of a pointer, \ie each \SI{64}{\bit}, if it lies in this region.
If so, we dump the contents of the page as well as all the offsets representing a pointer within the page.
We also extend this approach to kernel pages belonging to kernel modules, as they are also randomized in a certain memory region and could be used to break the randomization of the modules.
On a machine running Linux 5.4.92, we find \SIx{15737} pages where \SIx{4070} contain values matching pointers within these memory regions.

In a second step, we filter the dumped pages for possible candidates that we can use for the attack.
We try to find corresponding symbol names to the detected addresses by matching them to \texttt{/proc/kallsyms}, yielding \SIx{15} pages that only contain resolvable kernel text addresses.
\SIx{3973} pages contained module addresses, \SIx{39} resolvable and unresolvable addresses, and \SIx{43} no symbols at all.
These pages now need to be checked if their content is static and, thus, does not change over time.
This can be achieved by dumping the content periodically and checking it for modifications.
Further, we want the pages to not contain any data initialized during boot time and, thus, we need to check if the content of those pages changes (excluding the kernel addresses) while rebooting the system multiple times.
In order to rule out hardware-specific data, this should be done on different physical machines.

\subsubsection{Remote Attack}\label{sec:case-studies:kaslr:attack:remote}

For our remote KASLR break, we implemented the victim server in two ways.
First, as a RESTful API listening for HTTP/1 requests implemented in C++ using the pistache framework~\cite{Pistache}.
For simplicity reasons, the API allows the adversary to set and modify the content of pages directly.
However, as we have shown in~\cref{sec:case-studies:memcached}, the data could be stored in an in-memory database as well.
Second, we elevate the service for HTTP/2 to support multiplexing, allowing us to mount timeless timing attacks described by Van Goethem~\etal\cite{VanGoethem2020Timeless}.
In both scenarios, an Nginx~\cite{nginx} web server running on the target machine forwards the request to the victim service.
The attacker sends the crafted pages for the offset to test to the victim using the API.
After \SI{2}{\second}, \ie the time the page would be deduplicated on our system with a high chance, the attacker sends a network request modifying and, thus, causing the probable duplication, and measures its response time.

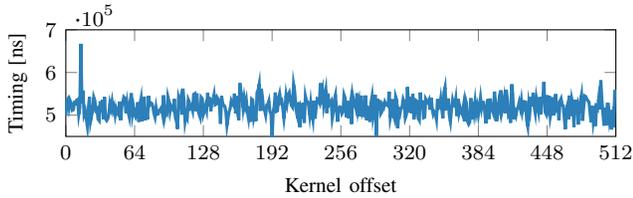
\begin{figure}[t!]
 \centering
 \tikzsetnextfilename{kaslr-remote1}
 \begin{tikzpicture}
\begin{axis}[
style={font=\footnotesize},
xlabel={Kernel offset},
ylabel={Timing [ns]},
y label style={align=center,text width=2cm},
width=\hsize,
xmin=0,
ymax=700000,
ymin=450000,
xmax=512,
height=3cm,
xtick={0,63,127,191,255,319,383,447,511},
xtick={0,64,128,192,256,320,384,448,512},
]

\addplot+[very thick,no marks] table[x=Offset,y=Value,col sep=comma] {data/kaslr-remote1.csv};
\end{axis}
\end{tikzpicture}
 \caption{Execution time to a page containing the content adjusted to one of the possible kernel offsets. 
 The high peak at offset 14 yielded the same content as the kernel page of the VM and, thus, has been deduplicated by the hypervisor.
 }
 \label{fig:case-studies:kaslr:attack:remote1}
\end{figure}

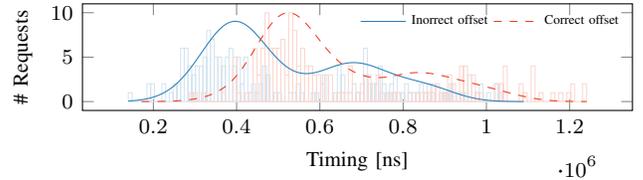
\begin{figure}[t!]
 \centering
 \tikzsetnextfilename{kaslr-hist}
 \begin{tikzpicture}
\begin{axis}[
style={font=\footnotesize},
xlabel={Timing [ns]},
ylabel={\# Requests},
y label style={align=center,text width=2cm},
width=\hsize,
height=3cm,
legend style={at={(1.0,1.0)}, anchor=north east, legend columns=2, font=\tiny,draw=none,fill=none},
]

\addplot[blue!20, fill=none, hist, hist/bins=100] table [y=Incorrect, col sep=comma] {data/kaslr_remote_hist.csv};
\addplot[red!20, pattern=north east lines, fill=none, hist, hist/bins=100] table [y=Correct, col sep=comma] {data/kaslr_remote_hist.csv};
\end{axis}

\begin{axis}[
style={font=\footnotesize},
xlabel={Timing [ns]},
ylabel={},
y label style={align=center,text width=2cm},
width=\hsize,
axis y line=none,
axis x line=none,
height=3cm,
legend style={at={(1.0,1.0)}, anchor=north east, legend columns=2, font=\tiny,draw=none,fill=none},
]

\addplot[blue, fill=none] table [x=IncorrectX, y=IncorrectY, col sep=comma] {data/kaslr_remote_hist_kde.csv};
\addlegendentry{Inorrect offset};
\addplot[dashed,red, fill=none] table [x=CorrectX, y=CorrectY, col sep=comma] {data/kaslr_remote_hist_kde.csv};
\addlegendentry{Correct offset};
\end{axis}

\end{tikzpicture}
 \caption{Histogram of the measured access times for an incorrect and the correct offset for the KASLR break. A correct guess can be clearly distinguished from an incorrect guess.}
 \label{fig:case-studies:kaslr:attack:hist}
\end{figure}

\paragraph{HTTP/1.}
In the first scenario, we use HTTP/1 to communicate with the network service and measure the response time of the network requests.
\cref{fig:case-studies:kaslr:attack:hist} illustrates the distribution of response times for the correct offset and an incorrect offset.
\cref{fig:case-studies:kaslr:attack:remote1} shows the mean response time of a network request for a specific offset in a remote-attack scenario.
After sending \SIx{100} requests, we can clearly see the increased response time for the currently used randomized kernel offset.

In the local setting, we were able to recover the correct randomization offset with a success rate of \SI{100}{\percent} and an average runtime of \SI{21.3}{\second} ($n=100$).
In the remote setting, we were able to recover the correct randomization offset with a success rate of \SI{73}{\percent} and an average runtime of \SI{5}{\minute} \SI{57.9}{\second} ($n=100$).
With the default configuration of the cloud provider (\cf \cref{sec:attack_primitives:memory_deduplication}) this would yield an average simulated runtime of \SI{34}{\minute} \SI{28.69}{\second}.
With the default Linux settings, it would take \SI{9} hours and \SI{2} minutes.

\paragraph{HTTP/2 Multiplexing.}
To improve on the measurement noise introduced by the connection between the victim and the adversary and the necessity of accurate time stamps, we utilize Timeless Timing attacks~\cite{VanGoethem2020Timeless} to overcome this issue.
HTTP/2 allows to pack multiple requests within a single packet and, thus, the requests reach the server at the same time.
However, the response of the request that reaches the sender faster has likely been processed quicker.

In contrast to the sequential HTTP/1 attack, we pick pairs of kernel offsets that we send to the server using multiplexed HTTP/2 requests.
For every attempt, we send each pair to the server and record for which request we receive the response first.
Note that we do not need to rely on measured access times but just on the response order of the requests.
For pairs of both incorrect kernel offsets, we should observe a uniform distribution between the offsets.
However, if one of the offsets is the correct one, we should observe an unequal distribution.
To optimize the approach, we reduce the number of candidates and filter out pairs with a uniform distribution early.
With each filter step, we re-combine the candidates to new pairs.

To amplify the signal, we crafted \SIx{7} kernel pages for each possible kernel offset.
In the local-network setting, we achieved a success rate of \SI{88.89}{\percent} with an average runtime of \SI{1} minute and \SI{38} seconds ($n=100$).
The raw timing differences observed in the HTTP/1 setting enable a faster attack than HTTP/2.
We were able to successfully find the correct offset in the remote setting with a success rate of \SI{92}{\percent} with an average runtime of \SI{3} minutes and \SI{15} seconds ($n=100$).
With a prolonged waiting time using the default configurations of the cloud provider of how many pages are scanned by the operating system per minute, this would yield an average simulated attack time of \SI{18} minutes and \SI{25} seconds.
With the default Linux settings, it would take \SI{4} hours and \SI{48} minutes.

\subsection{InnoDB Record Data Leakage}\label{sec:case-studies:mariadb}
InnoDB is a storage engine used by default in the database management systems MySQL and MariaDB.
The storage engine efficiently buffers record data and index caches in the memory and is used instead of using the operating system's page cache directly.
InnoDB has the advantage of providing faster access to frequently used data.

Database systems use indices to allow quick access to records, \ie normally, an index is placed automatically on columns marked as primary key.
InnoDB implements indices using a B+ tree, which allows fast record lookups.
The nodes of the tree are represented by index pages, which are the basic storage unit of InnoDB and have a size of \SI{16}{\kilo\byte} by default.
The leaf index pages contain the actual user data.
The non-leaf ones link to other leaf or non-leaf pages.
Index pages on the same tree level are linked together to allow scanning operations.
User records in an index page are logically linked in ascending order by their key but may be placed anywhere in the page's physical memory.

\paragraph{High-Level Overview of the Attack.}
To achieve byte-by-byte leakage, the attacker needs to control the content and size of data that is stored before the target data to bytewise shift the target data onto the attacker-controlled page.
Bosman~\etal\cite{Bosman2016} showed that byte-by-byte leakage is possible.

InnoDB performs a data reorganization of data if an insert or update query fails as an optimization.
This optimization enables byte-by-byte leakage if the attacker controls most of the InnoDB record.
Using this primitive to perform memory massaging, an attacker can shift the secret.

\paragraph{Assumptions.}
We assume that Memcached can be used in addition as a leakage primitive to leak the secret data co-located to the attacker-controlled data bytewise.
As we will later analyze, the Linux page cache caches 
Note that this can be any primitive used for triggering deduplication and copy-on-write page faults, \ie nginx, as was shown by Bosman~\etal\cite{Bosman2016}.
We assume a database application with a user table which is defined in~\cref{fig:user_table} and that the InnoDB index page has a certain layout, which is explained in more detail in~\Cref{sec:attack_analysis}.
We assume that the attacker can perform multiple tries in parallel until such a layout is given.
If the layout is given, the attacker can verify whether the requirements are fulfilled.

\begin{figure}[t]
  \centering
   \resizebox{\hsize}{!}{
     \tikzsetnextfilename{innodb-high-level}
     \input{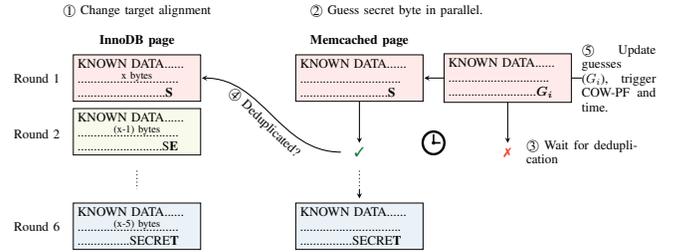}
   }
 \caption{High-level idea of the InnoDB Reorganization attack.}
 \label{fig:innodb_high_level}
 \end{figure}

\paragraph{Attack steps.}
\Cref{fig:innodb_high_level} illustrates the five steps of the InnoDB reorganization attack.
In the first round, the attacker triggers the reorganization and shifts the first byte of the secret value (\textbf{``SECRET''}) onto the controlled \SI{4}{\kilo\byte}-page.
Next, the attacker stores multiple guesses into Memcached. 
The attacker waits until the deduplication happened. 
After the deduplication happened, the attacker updates the Memcached guess pages and measures the round-trip time of the network packets.
The right guess should lead to a significantly higher timing than the other guesses with enough samples taken.
Afterwards, the attacker repeats the procedure to shift the second byte into the attacker-controlled InnoDB page, updates the guesses in Memcached, including the first recovered byte, and leaks the second byte.
This procedure can be repeated up to a certain leakage size. 
The limits are discussed in~\Cref{sec:appendix-furtherreqs}.

\paragraph{Why an Additional Leakage Primitive is Required.}
InnoDB tries to circumvent the page cache of the Linux kernel by using the \texttt{O\_DIRECT} flag in \texttt{mmap}~\cite{InnoDBDataFlushing}.
However, the data is still in the page cache and gets deduplicated.
The page-cached data cannot be overwritten directly via InnoDB.
Therefore, we cannot use a second InnoDB record to trigger a copy-on-write page fault since the data would also get deduplicated.
We found no convenient and reliable way to get external blobs consistently in the memory and replace them to trigger copy-on-write page faults in InnoDB.
For external blobs, we have a similar race as in Memcached, since updates are not performed in-place.
Instead, resource releasing is performed in a similar way compared to Memcached.
Consequently, for our attack, we use a memory-resident second channel (Memcached) to trigger the copy-on-write page fault.
However, this could, in general, be any web application/resource providing such a leakage primitive that is running on the same machine.

\subsubsection{Determining InnoDB Attack Requirements.}
We analyze the memory ordering of InnoDB by performing different SQL statements:

\paragraph{Insert.}
Upon inserting a new record, InnoDB first tries to place the record in its corresponding index page.
If no such page exists, a new one is created.

In case of an existing index page with sufficient consecutive free space, \eg unused space at the end of the page or a gap from previous deletes, the record is placed on this index page.
Should this not be possible, \eg the page is full, or the free space is too fragmented, either the current index page is defragmented (\emph{reorganized}), a new page is allocated, or a page \emph{split} is performed.
Inserting into a new index page is only possible if it does not break the existing relations in the index tree.
Otherwise, a page split has to be performed.
As the space of previously deleted records is reused, the physical order of records in an index page does not always reflect their logical order, \eg a record with key 5 might be inserted in memory before a record with key 2.  

\paragraph{Delete.}
When a record is deleted, it is added to the index pages free record list. 
Should the free space resulting from deletions reach a certain merge threshold, InnoDB tries to  perform a \texttt{merge} operation to save space.
A merge operation is possible if the utilization of the next or previous linked index page is low enough to combine it with the current page~\cite{InnoDBPageMergingandSplitting}.

\paragraph{Update.}
Update queries in InnoDB update a record in-place, as long as the updated record fits in the same size as the old one (\texttt{new\_record\_size} $\leq$ \texttt{old\_record\_size})~\cite{InnoDBRowUpdate}.
Otherwise, the update operation is realized as a delete with a subsequent insert operation, inserting the updated record.

\paragraph{Reorganization.}
An insert or update query can fail even if enough space is available on the index page because the free space is fragmented.
In such a case, InnoDB performs an optimization called \texttt{reorganization}~\cite{InnoDBReorganize}. 
During reorganization, the page is rebuilt by clearing its contents and inserting existing records in their logical order.
Afterwards, the pending insert or update operation is completed using the freed space at the end of the page.

\subsubsection{Reorganization Attack.}
As shown by Bosman~\etal\cite{Bosman2016}, an attacker can use memory-deduplication attacks to leak data byte-by-byte if the attacker can change the memory layout on a byte granularity.
We demonstrate that this approach can also be applied in a fully remote attack scenario.

We leak data byte-by-byte by exploiting the \texttt{reorganization} of database records in InnoDB index pages.
The reorganization is triggered if data is updated or inserted, and reorganization keeps the data on the same index page.
\Cref{fig:user_table} in~\cref{sec:appendix:mariadb} shows the user table and its fields in the database, including an \texttt{id}, \texttt{username}, \texttt{password}, and an \texttt{image} field. 
We assume that the attacker can register an arbitrary number of users and modify their content.

\paragraph{Alignment Changing.}
To leak attacker-unknown record data, we need a large record $r_{AT}$ to \texttt{shift} bytes from a target record $r_{T}$ into an attacker-controlled region.
To trigger the \texttt{reorganization}, we require an additional record $r_{AX}$ in the user table.
The reorganization orders the records in RAM in their logical order.
With targeted size modifications of the attacker-controlled records $r_{AT}$ and $r_{AX}$, we can trigger the reorganization and bytewise shift record data from $r_{T}$ into an attacker-controlled \SI{4}{\kilo\byte} region.
To leak the targeted byte, we use Memcached used in a simple HTTP web server as a second channel, same as in~\cref{sec:case-studies:memcached}.

\begin{figure}[t]
  \centering
   \resizebox{\hsize}{!}{
     \tikzsetnextfilename{amplification-byte}
     \input{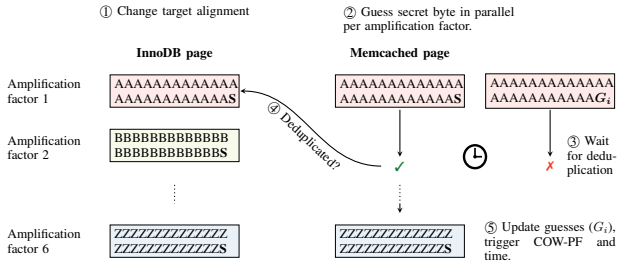}
   }
 \caption{Leakage of a secret byte (\textbf{S}) from an InnoDB record using Memcached with amplification.}
 \label{fig:ampl_memcached}
 \end{figure}

\paragraph{Amplification.}
To use amplification, we fill multiple pages on both Memcached and InnoBD with different fill bytes but with a constant leaked offset, as shown in~\cref{fig:ampl_memcached}.
As already mentioned, the copy-on-write bit stays set for the correct guess in Memcached.
Therefore we can amplify by updating both Memcached and InnoDB fill bytes and waiting again for the deduplication.
This procedure can be repeated up to a certain amplification factor.
To trigger copy-on-write pagefaults, we send an HTTP request to the server, which overwrites the content of the Memcached pages.
To reset to an index page layout, which allows leaking a different byte offset, it is required to trigger another reorganization by modifying the sizes of the records $r_{AX}$ and $r_{AT}$.
All requirements are explained in full detail in~\cref{sec:attack_analysis}.

\subsubsection{In Detail Analysis of Attack Requirements.}\label{sec:attack_analysis}
To perform the reorganization attack, the attacker and victim have to be placed on the \textbf{same} InnoDB index page.
While this is a question on the workload of the system, we assume that an attacker can perform sufficient repetitions to generate a layout leading to data leakage.

A \texttt{reorganization} can be caused by updating the size of a record so that it does initially not fit in any available consecutive free space on the page but does fit after defragmentation.
By choosing record sizes in the right way, it can be guaranteed that such reorganizations are always possible.

\begin{figure}[t]
     \begin{subfigure}{\hsize}%
        \centering
        \resizebox{0.8\hsize}{!}
        {
            \tikzsetnextfilename{innodb-record}
            \begin{tikzpicture}[]

\draw[fill=green!10] (0,0) rectangle +(1.5,-3.25) node [pos=.5] {\parbox{3.5cm}{\centering \textbf{header}}};
\draw[pattern=north west lines] (1.5,0) rectangle +(1.5,-3.25) node [pos=.5] {};
\draw[fill=red!10] (3,0) rectangle +(2.5,-3.25) node [pos=.5] {\parbox{3.5cm}{\centering \textbf{$r_{AX}$}}};
\draw[pattern=north west lines] (5.5,0) rectangle +(1.5,-3.25) node [pos=.5] {};
\draw[fill=red!10] (7,0) rectangle +(1.5,-3.25) node [pos=.5] {\parbox{3.5cm}{\centering \textbf{$r_{AT}$}}};
\draw[fill=blue!10] (8.5,0) rectangle +(1,-3.25) node [pos=.5] {\parbox{3.5cm}{\centering \textbf{$r_{T}$}}};
\draw[] (9.5,0) rectangle +(1,-3.25) node [pos=.5] {};
\draw[fill=green!10] (10.5,0) rectangle +(1.5,-3.25) node [pos=.5] {\parbox{3.5cm}{\centering \textbf{footer}}};

\node at (6,.5) {\textbf{InnoDB Index Page}};
\node at (6,-4) {\textbf{Initial State}};

\node at (0,-4) {\textbf{0 kB}};
\node at (12,-4) {\textbf{16 kB}};

\node[] at (2.25,.75) {\footnotesize Record data};
\node[rotate=90] at (10,-1.5) {\footnotesize trailing free space};

\draw[->,>=stealth,thick] (2.25,.5) to node[midway,above] {}(2.25,0);

\end{tikzpicture}%
         }
         \caption{InnoDB page layout, which is susceptible to a reorganization attack. A simple model of an index page consists of a fixed-size header, user records, the unused space at the page end, and a footer. In this scenario the attacker controls the records $r_{AX}$ and $r_{AT}$. $r_{AT}$ is used to control the alignment of target record to leak $r_{T}$. $r_{AX}$ is used to trigger the reorganization within an InnoDB index page.}
         \label{fig:innodb_page}
     \end{subfigure}

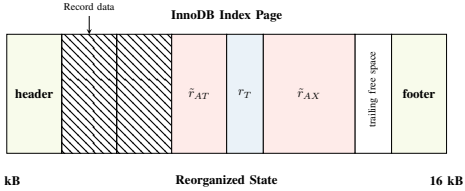
     \begin{subfigure}{\hsize}%
         \centering
         \resizebox{0.8\hsize}{!}
         {
            \tikzsetnextfilename{innodb-record-reorganized}
            \begin{tikzpicture}[]

\draw[fill=green!10] (0,0) rectangle +(1.5,-3.25) node [pos=.5] {\parbox{3.5cm}{\centering \textbf{header}}};
\draw[pattern=north west lines] (1.5,0) rectangle +(1.5,-3.25) node [pos=.5] {};
\draw[pattern=north west lines] (3,0) rectangle +(1.5,-3.25) node [pos=.5] {};
\draw[fill=red!10] (4.5,0) rectangle +(1.5,-3.25) node [pos=.5] {\parbox{3.5cm}{\centering \textbf{$\tilde{r}_{AT}$}}};
\draw[fill=blue!10] (6,0) rectangle +(1,-3.25) node [pos=.5] {\parbox{3.5cm}{\centering \textbf{$r_{T}$}}};
\draw[] (9.5,0) rectangle +(1,-3.25) node [pos=.5] {};
\draw[fill=red!10] (7,0) rectangle +(2.5,-3.25) node [pos=.5] {\parbox{3.5cm}{\centering \textbf{$\tilde{r}_{AX}$}}};
\draw[fill=green!10] (10.5,0) rectangle +(1.5,-3.25) node [pos=.5] {\parbox{3.5cm}{\centering \textbf{footer}}};

\node at (6,.5) {\textbf{InnoDB Index Page}};
\node at (6,-4) {\textbf{Reorganized State}};

\node at (0,-4) {\textbf{0 kB}};
\node at (12,-4) {\textbf{16 kB}};

\node[] at (2.25,.75) {\footnotesize Record data};
\node[rotate=90] at (10,-1.5) {\footnotesize trailing free space};

\draw[->,>=stealth,thick] (2.25,.5) to node[midway,above] {}(2.25,0);

\end{tikzpicture}%
         }
         \caption{ If the reorganization was triggered $\tilde{r}_{AX}$ is moved to the beginning of the trailing free space.}
         \label{fig:innodb_reorganization}
     \end{subfigure}
     
     \begin{subfigure}{\hsize}%
         \centering
         \resizebox{0.8\hsize}{!}
         {
            \tikzsetnextfilename{innodb-record-reset}
             \begin{tikzpicture}[]

\draw[fill=green!10] (0,0) rectangle +(1.5,-3.25) node [pos=.5] {\parbox{3.5cm}{\centering \textbf{header}}};
\draw[pattern=north west lines] (1.5,0) rectangle +(1.5,-3.25) node [pos=.5] {};
\draw[pattern=north west lines] (5.5,0) rectangle +(1.5,-3.25) node [pos=.5] {};
\draw[fill=red!10] (3,0) rectangle +(2.5,-3.25) node [pos=.5] {\parbox{3.5cm}{\centering \textbf{$r_{AX}$}}};
\draw[fill=blue!10] (7,0) rectangle +(1,-3.25) node [pos=.5] {\parbox{3.5cm}{\centering \textbf{$r_{T}$}}};
\draw[] (9.5,0) rectangle +(1,-3.25) node [pos=.5] {};
\draw[fill=red!10] (8,0) rectangle +(1.5,-3.25) node [pos=.5] {\parbox{3.5cm}{\centering \textbf{${r}_{AT}$}}};
\draw[fill=green!10] (10.5,0) rectangle +(1.5,-3.25) node [pos=.5] {\parbox{3.5cm}{\centering \textbf{footer}}};

\node at (6,.5) {\textbf{InnoDB Index Page}};
\node at (6,-4) {\textbf{Reset State}};

\node at (0,-4) {\textbf{0 kB}};
\node at (12,-4) {\textbf{16 kB}};

\node[] at (2.25,.75) {\footnotesize Record data};
\node[rotate=90] at (10,-1.5) {\footnotesize trailing free space};

\draw[->,>=stealth,thick] (2.25,.5) to node[midway,above] {}(2.25,0);

\end{tikzpicture}%
         }
         \caption{Reset and reorganized InnoDB record.}
         \label{fig:innodb_reset}
     \end{subfigure}
 \caption{InnoDB reorganization steps.}
 \label{fig:record_structure}
\end{figure}
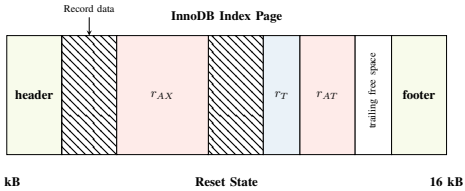

\paragraph{Initial Page Layout.}
\cref{fig:innodb_page} describes the initial page layout required by InnoDB to leak record data.
We exploit InnoDB's reorganization feature as a primitive for the attacker to align the secret on a byte granularity.
At the beginning of an index page, there are a couple of headers and system records, summing up to \SI{120}{\byte} ~\cite{InnoDBPage0PageIC}.
Then the data of the user records follows.
At the end of the page, there is a so-called page directory and further meta-data.
The user records are also preceded by a dynamic-sized header, which depends on the table layout and contains information necessary for using and organizing the records.
\cref{fig:innodb_page} shows the assumed initial physical and logical layout for our InnoDB attack.
The hatched areas represent unknown records.
$r_{AT}$ and $r_{AX}$ are attacker-controlled records and $r_{T}$ is the target record to leak.

\paragraph{Analysis of Required Sizes for Exploiting Reorganization.}
During the rebuilding of index pages, records are inserted consecutively in memory by their logical order, except for the record that triggered the reorganization, which is inserted last, regardless of its key.
In total, it is possible to insert \SI{16252}{\byte} (\texttt{max\_free\_space}) of record data into an index page.
The layout requires that the record $r_{AT}$ is logically located before $r_{T}$. 
$r_{AX}$ is required to be physically before the two records, somewhere in between the two hatched regions in~\cref{fig:innodb_page}.
The record $r_{AT}$ is used to change and control the target record's alignment $r_{T}$.
The attacker wants to make $r_{AT}$ as large as possible to change the target record's $r_{T}$ alignment.
In the default setting of InnoDB with a default index page size of \SI{16}{\kilo\byte}, the maximum size for a record is \SI{8125}{\byte}~\cite{InnoDBPage0PageIC}.
Thus, to leak as much data as possible, we choose $r_{AT}$ to be \SIx{8125}.
The validation of all requirements and the potential leakage rate is described in~\cref{sec:appendix-furtherreqs}.
Next, we discuss the attack steps in more detail.

\paragraph{Preparing the Alignment and Triggering the Reorganization.}
To trigger a reorganization, the attacker increases the size of record $r_{AX}$ using an update query.
The reorganization only happens if the updated size still fits into the total free size of the index page.
The new reorganization moves $r_{AX}$ to the \texttt{trailing free space} within the index page, which is large enough to contain at least $r_{AX} + 1$.

If the attacker wants to shift a byte of the target record by $\delta$ bytes such that the byte moves closer to $r_{AT}$, the attacker can update the size of $r_{AT}$ and decrease it by $\delta$ and increase the size of $r_{AX}$ by $\delta$.
The reorganization takes out the record $r_{AX}$ and moves it to the newly created free location.
It causes the record $\tilde{r}_{AX}$ to be moved after $\tilde{r}_{AT}$ and $r_{T}$.
$r_{AX}$ can only be modified up to the maximum record size.
While the header of the user record has a dynamic size, we assume that the record does not change during the attack.
If there is an additional record after $r_{T}$, the header stays the same.
If there is no record after $r_{T}$, the record header points to the Supremum~\cite{InnoDBInfimumSupremum}, which is at the beginning of the page.
With each alignment change, the next offset field in the records header needs to be incremented by the byte offset to leak.

\paragraph{Leaking the Secret Byte.}
Since we cannot leak the records in MariaDB alone on Linux, a second way to trigger the memory deduplication is required.
Furthermore, we apply amplification for our bytewise leakage, as discussed in~\cref{sec:remote_covert_channel}.
\cref{fig:ampl_memcached} illustrates the amplified version of the InnoDB record attack.
We use different pages with the same content but only a different last byte in the page, which is our probe byte in Memcached (in~\cref{fig:ampl_memcached} the secret byte is $x$).
In MariaDB, we update our record $r_{AT}$ with the content of the first page (\texttt{AAA....S}).
Afterwards, we wait for a certain delay until the memory deduplication is triggered.
If the secret byte is correct, the page gets deduplicated.
After the delay, we modify $r_{AT}$ again.
The page in Memcached still has the R/W bit cleared.
If all amplification pages are deduplicated, we use the web application to write on each of our amplification pages and measure the response time.

\paragraph{Reset.}
After the reorganization, the alignment is changed, and we get a record layout, as illustrated in~\cref{fig:innodb_reorganization}.
Unfortunately, we cannot modify the base alignment since we do not know the size of the other records on the index page.
However, we can either try to repeat the attack until we start at the beginning of a \SI{4}{\kilo\byte} page or leak the base alignment. 
To reset the state back to the initial one, we again exploit reorganization, changing to the previous sizes.
The requirement to trigger another reorganization via $r_{AT}$ is that the trailing free space is smaller than the reset size of $r_{AT}$.
The reorganization causes that the updated record $r_{AT}$ is now moved after $r_{T}$, leading to the memory layout illustrated in~\cref{fig:innodb_reset}.
However, this is no problem since we can force another reorganization, bringing back our reorganized state, as illustrated in~\cref{fig:innodb_reorganization}.
After each alignment change, we switch between the reset and reorganized state and never return to the initial state.

\paragraph{Evaluation.}
We implement and evaluate our attack on MariaDB version 10.5.8, using UNIX sockets and a simple HTTP server to connect to it and to Memcached 1.6.8.
The database is setup as shown in~\cref{fig:innodb_page}.
We choose a random secret of \SI{4}{\byte} and repeat our data leakage experiment \SIx{20} times.
We apply the amplification technique shown to leak a single byte via \SIx{8} pages. 
To be on the safe side, we send \SIx{40} requests for all 256 possibilities.
Afterwards, we probe all \SIx{256} possibilities for the secret byte at once via Memcached.
We look at the timing difference between the means of the received distribution of writing to copy-on-write and non-copy-on-write pages.
Our attack automatically detects if a byte was accidentally classified as copy-on-write in case we do not get clear results for the following byte to leak. 
In this case, we can backtrack to the last byte that was correctly guessed.
Therefore, our approach is self-correcting in case we accidentally received a wrong byte, and, thus, our approach is nearly complete error-free, despite the last byte where an error might occur.

On Linux, we observed that the time to wait for the deduplication on InnoDB is, in many cases, more than twice as big as in the previous cases.
To be on the safe side, we increased the wait time to \SIx{4} seconds.
As the amplification needs to be triggered sequentially, this leads to a wait time of \SIx{32} seconds per guess round.
This longer delay is required since the target page is constantly changed, and KSM does not immediately deduplicate pages which are often modified~\cite{RedHatKSM}.
The runtime of the attack to leak four random bytes is on average \SIx{5644.20} seconds ($n=100,\sigma=0.54\%$).
Thus, the attack leaks on average a single byte in \SIx{39.07} minutes or about \SI{1.5}{\byte/\hour} from a virtual machine running on a remote server in the local area network.
We simulate the attack's performanc using the default configuration to \SI{0.018}{\byte/\hour}.
With the provided configuration of the cloud provider, we got a simulated time of \SI{0.07}{\byte/\hour}.
Note that the large bottleneck of this attack is the amplification technique \ie for one iteration \SI{32}{\second} have to be waited.

\paragraph{Limitations.}
For the initial setup, \cf~\cref{fig:innodb_page}, the uncontrolled record data before $r_{AT}$ can be modified in-place as long as the overall size is not changed.
Every in-place update of other records does not influence the attack.
However, a memory split, merge, or reorganization would interfere with the attack and potentially destroy the needed layout.

\begin{tcolorbox}[title=C3]
\textbf{Find remote request paths that do not only keep attacker-controlled data in memory but also provide the attacker with control over alignment and in-memory representation.}
\tcblower
We demonstrated a scenario for InnoDB used in MariaDB and MySQL, which allows changing the alignment of database records remotely.
By changing the sizes of two attacker-controlled records, an attacker can load bytewise parts of victim's data to an attacker-controlled \SI{4}{\kilo\byte} page.
Amplification can be achieved by leveraging the fact that deduplication can be triggered multiple times by modifying the attacker-controlled record and adding certain amplification pages to Memcached (like shown in~\Cref{fig:ampl_memcached}.)
\end{tcolorbox}

\section{Mitigations and Further Attack Targets.}\label{sec:mitigations}

\subsection{Mitigations}
Our attack showed that memory deduplication is still a threat and even exploitable over the network. 
Even isolation into security domains like performed on Windows is not enough to mitigate information disclosure via memory deduplication.

\paragraph{Deactivation.}
While the simplest solution would be to altogether disable memory deduplication on Windows and Linux (Ubuntu), it is probably the most costly in terms of performance overhead.
Especially on Windows server, where multiple users would use the same application, this could lead to immense memory overhead.
Windows allows disabling of memory deduplication per process~\cite{WindowsProcessMitigationSideChannel}.

\paragraph{Only Deduplicate Zero Pages.}
Another mitigation by Bosman~\etal\cite{Bosman2016} would be only to deduplicate zero pages. 
According to their evaluation, between 84\% and 94\% of the deduplication in Microsoft Edge are only zero pages~\cite{Bosman2016}.
However, the covert channel is still possible with this solution since we can still trigger copy-on-write page faults on deduplicated zero pages.

\paragraph{TPS.}
VMWare TPS~\cite{Vmware2021TransparentPageSharing} uses additional salts to enable memory deduplication. 
The salt value and the content of page have to be identical to be shared.
If VMs want to deduplicate shared content, the salted value is unknown to an attacker.
While this approach protects against cross-VM attacks, TPS does not protect against remote memory-deduplication attacks in the same domain.

\paragraph{CovertInspector.}
Wang~\etal\cite{Wang2015Covert} demonstrated an approach to detect memory-deduplication attacks by modifying KVM by 300 lines of code.
Their approach has a particular focus on intercepting the \texttt{rdtsc} instruction triggered by the VM and also the number of pagefaults.
Remote timers are not considered by CovertInspector.

\paragraph{VUsion.}
VUsion~\cite{Oliverio2017Secure} mitigates all kinds of memory-deduplication attacks by applying a share-XOR-fetch policy and fake merging.
All pages that are considered for deduplication behave the same in terms of access times and copy-on-write pagefaults.
Fake merging guarantees that every access on a page, both shared or non-shared, behaves the same in terms of access time.
This mechanism prevents attacks on the detection of pages being actually deduplicated~\cite{Oliverio2017Secure}.
While fake merging would mitigate all of our attacks based on the copy-on-write page fault, it is not implemented nor intended to be merged in the Linux kernel.

\begin{figure}[t!]
 \centering
 \tikzsetnextfilename{vusion}
 \begin{tikzpicture}
\begin{axis}[
style={font=\footnotesize},
xlabel={Kernel offset [MB]},
ylabel={Execution Time},
y label style={align=center,text width=1.5cm},
width=0.9\hsize,
xmin=1,
xmax=512,
ymin=50000,
ymax=200000,
height=3cm,
xtick={0,63,127,191,255,319,383,447,511},
xtick={0,64,128,192,256,320,384,448,512},
legend style={at={(0.0,1.0)}, anchor=north west, legend columns=2, font=\tiny,draw=none,fill=none},
]

\addplot+[thick,no marks,red!50,densely dotted] table[x=Offset,y=With,col sep=comma] {data/vusion.csv};
\addlegendentry{Vusion};
\addplot+[thick,no marks,blue] table[x=Offset,y=Without,col sep=comma] {data/vusion.csv};
\addlegendentry{Without Vusion};

\end{axis}
\end{tikzpicture}
 \caption{Mean response time for all possible kernel offsets. While an adversary can easily observe the correct offset \SIx{106} on an unprotected system (blue), the VUsion-protected system (red) prevents the leakage.}
 \label{fig:mitigations:vusion}
\end{figure}
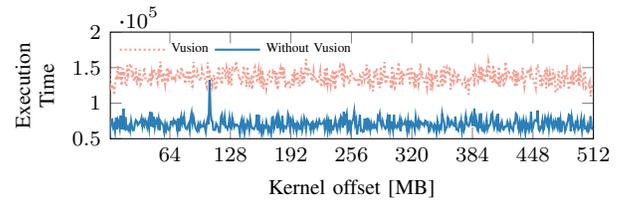

We experimentally verified the effectiveness of VUsion against our remote memory-deduplication attacks in a local area network setting.
\cref{fig:mitigations:vusion} illustrates the KASLR break (\cf \cref{sec:case-studies:kaslr}) on a protected (red) and an unprotected Linux kernel 4.10 (blue) running Ubuntu 17.04 LTS.
We measured the response time for every possible offset 100 times and reported the mean value.
One can clearly see that our attack successfully recovers the correct offset \SIx{106} while the attack against the VUsion-protected kernel only observes higher timings.

\paragraph{Network-layer Countermeasures.}
On the network layer, we can mitigate remote memory-deduplication attacks via network packet inspection tools and DDoS monitoring. 
Another possibility to mitigate remote memory-deduplication attacks is by adding additional noise to the network, \ie by performing load balancing or adding discrete time delays.
This would require more samples for the attacker, and at a certain point, it could make the attack infeasible.

\subsection{Alternative Attack Targets}
We want to emphasize that fixing Memcached does not mitigate the problem of remote memory-deduplication attacks as our techniques are generic and can be applied to other applications as well.
In addition to Memcached and InnoDB, we analyzed further applications which could be susceptible to remote memory-deduplication attacks.
Many web applications offer the possibility to use Memcached, such as PHPBB, WordPress, Moodle, and PrestaShop. 
Moodle allows image caching, which might be already used to perform the fingerprinting attack.
We analyzed the in-memory DB Redis and found that \SI{4}{\kilo\byte} pages can be also placed into the memory.
There is again meta-data stored about the stored item, and it is again a question of the correct alignment for the attacker to perform remote memory-deduplication attacks.
If the attacker's guess about the alignment is correct, copy-on-write pagefaults can be triggered in a similar manner to Memcached by freeing an item and again inserting a new one with the equal size. 
This leads to an overwrite of the deduplicated memory.
Furthermore, we analyzed the other popular alternative for in-memory databases SQLite. 
However, we found that we could not fully place a single \SI{4}{\kilo\byte} page into memory.
We also checked Aerospike and observed that memory is in DRAM as key-value pair and that the \texttt{aerospike\_key\_put} function directly replaces the content and could be used to trigger copy-on-write pagefaults.
As already shown by Bosman~\etal\cite{Bosman2016} also request pools like used in nginx are susceptible to memory deduplication attacks.

\section{Conclusion}\label{sec:conclusion}
In this work, we presented how memory deduplication can be exploited from a remote perspective.
This attack does neither require local code execution nor JavaScript execution in the browser, as demonstrated in previous work.
With targeted web requests, we can observe timing differences between duplicated pages over the network.
We first evaluated the speed of our remote covert channel based on an HTTP web server achieving a performance of up to \SI{302.16}{\byte/\hour} in a LAN setting and \SI{34.41}{\byte/\hour} over the internet.
Further, we fingerprinted libraries used on the system by exploiting the Memcached database.
It is possible to fingerprint libraries within \SI{166.51}{\second} over the internet.
Within only \SIx{4} minutes, we successfully broke KASLR from a virtual machine running on a server 14 network hops away.
Even though there are potential mitigations against memory deduplication within the same security domain, they are not applied in Linux systems.
Finally, we leaked the database records' content from InnoDB with \SI{1.5}{\byte/\hour}.

\section*{Acknowledgments}
We would like to thank our anonymous reviewers for valueable feedback and comments on the paper.
Furthermore, we want to thank Tom Van Goethem for feedback on the draft and support on the HTTP/2 experiments.
We want to thank Equinix Metal for providing us bare metal servers.
This work was supported by generous funding and gifts from the EU project SOPHIA, Red Hat and AWS.
Any opinions or recommendationsw expressed in this work are those of the authors and do not necessarily reflect the views of the funding parties.

\bibliographystyle{plainurl}
\bibliography{main}

\appendix

\section*{Timing Difference of Library Fingerprinting in PHP}\label{sec:appendix-timingdiffphp} 
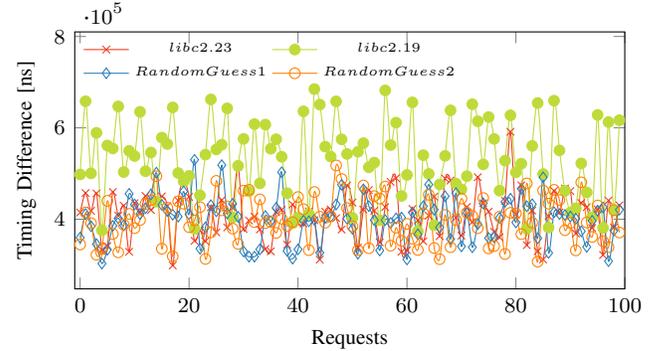
\begin{figure}[h]
    \centering
    \tikzsetnextfilename{php_timing_fingerprinting}
    \begin{tikzpicture}
            \begin{axis}[
            enlarge x limits={0.01},
            style={font=\footnotesize},
            width=\hsize,
            height=5cm,
            xlabel={Requests},
            ylabel={Timing Difference [ns]},
            ymax=810000,
            legend style={at={(0,1)}, anchor=north west, legend columns=2, font=\tiny,draw=none,fill=none},
            grid style=dashed
            ]

            \addplot[smooth,red,mark=x]  table[x index={0},y index = {1}, col sep=comma]{data/php_timestamps_stacked.csv};
            \addlegendentry{$libc 2.23$}

            \addplot[smooth,green,mark=*]  table[x index={0},y index = {2}, col sep=comma]{data/php_timestamps_stacked.csv};
            \addlegendentry{$libc 2.19$}

            \addplot[smooth,blue,mark=diamond]  table[x index={0},y index = {3}, col sep=comma]{data/php_timestamps_stacked.csv};
            \addlegendentry{$Random Guess 1$}

            \addplot[smooth,orange,mark=o]  table[x index={0},y index = {4}, col sep=comma]{data/php_timestamps_stacked.csv};
            \addlegendentry{$Random Guess 2$}

            \end{axis}
\end{tikzpicture}
    \caption{Timing difference for mapped libc version (2.19) vs. other guesses.}
    \label{fig:php_timing_fingerprinting}
\end{figure}

The copy-on-write page faults can be observed in PHP when triggering the deduplication via Memcached~\Cref{fig:php_timing_fingerprinting}.

\section*{Validation of Requirements for Reorganization.}\label{sec:appendix-furtherreqs} 
For mounting a successful oracle attack against an InnoDB record, it has to be guaranteed that a reorganization can be triggered reliably. 
Reorganizing is needed to switch between the different states introduced in Section~\ref{sec:attack_analysis}.

The condition for the first reorganize from the initial state (Figure~\ref{fig:innodb_page}) to the reorganized state (Figure~\ref{fig:innodb_reorganization}) is already guaranteed by the calculation of the initial size of  $|r_{AX}|$ in Section~\ref{sec:attack_analysis}.

For the switch from the reorganized state in Figure~\ref{fig:innodb_reorganization} to the reset state in Figure~\ref{fig:innodb_reset} it must be guaranteed that the restoring of $|r_{AT}|$ always triggers reorganization. Therefore the following inequality must hold:
{\footnotesize%
\begin{align*}
|r_{AT}| > max\_free\_space - |\tilde{r}_{AT}| - |r_{T}| - |r_{AX}| - footer\_sz
\end{align*}
We can claim that $|\tilde{r}_{AT}| + |r_{T}| \geq 4096$ must hold as otherwise the attacker does not even control one full page which is needed for the deduplication side channel. Using this and neglecting the \texttt{footer\_sz} we get:
\begin{align*} 
|r_{AT}| = 8125 &> max\_free\_space - (|\tilde{r}_{AT}| + |r_{T}|) - |r_{AX}| \\
                 &> 16252 - 4096 - 4064 = 8092
\end{align*}
}%
For the last state switch from the reset state to a new reorganized state there are two possibilities: $|r_{AX}|$ is either increased by $\delta$ as long as the resulting size is smaller than the maximum record size or it is set to the maximum record size. In both cases a reorganize should be triggered.
Therefore for case 1 the following inequality must hold:
{\footnotesize%
\begin{align*}
|r_{AX}| + \delta &> max\_free\_space - (|r_{AT}| - \delta) - |r_{T}| - |r_{AX}| \\
&- footer\_sz \\
2 * |r_{AX}| &> max\_free\_space - |r_{AT}| \\
2 * 4064 = 8128 &> 16252 - 8125 = 8127
\end{align*}
}%
For case two we again use that $|\tilde{r}_{AT}| + |r_{T}| \geq 4096$ must hold:
{\footnotesize%
\begin{align*}
|r_{AX,max}| &> max\_free\_space - |\tilde{r}_{AT}| - |r_{T}| - |r_{AX}| - footer\_sz \\
|r_{AX,max}| &> max\_free\_space - (|\tilde{r}_{AT}| + |r_{T}|) - |r_{AX}| \\
8125 &> 16252 - 4096 - 4064 = 8092
\end{align*}
}%

\paragraph{Required Sizes of Records and Potential Leakage Rate.}
The record $r_{AX}$ is required to trigger the reorganization of records.
Therefore, it initially has to be large enough so that we can trigger a reorganization.
We determine the worst case size of the left free space for records within an index page as follows after the first reorganization:

{\footnotesize
\begin{myalign*}
|r_{AX}| + 1  &> \textnormal{trailing free space, which is always the case if} \\
|r_{AX}| + 1 &> \texttt{max\_free\_space} - |r_{AX}| - |r_{AT}| - |r_{T}| (-\texttt{footer\_sz}). \\ 
&\textnormal{\texttt{footer\_sz}, $|r_{T}|$ can be neglected in worst case inspection} \\
|r_{AX}| &> \frac{\texttt{max\_free\_space} - |r_{AT}|-1}{2} = \SI{4064}{\byte} \\
\textnormal{ therefore:} \\
&\texttt{left\_free\_space} = \SI{16252}{\byte} - \SI{8125}{\byte} - \SI{4064}{\byte} = \SI{4063}{\byte}.
\end{myalign*}
}%

Next we want to determine the boundaries for the shift into our attacker-controlled \SI{4}{\kilo\byte}-page and the requirements.
We define the maximum alignment change $max\_alignment\_change$ as $r_{AT} - r_{AT_{header}}$.
To leak data from $r_{T}$, one page of our attacker-controlled $\tilde{r}_{AT}$ record needs to be page aligned.
As we chose the size of ${r}_{AT}$ to be \SI{8125}{\byte}, we do not fully control \SIx{2} pages.
We use a certain part of ${r}_{AT}$ to bring the last \SI{4096}{\byte} into a page alignment.
A certain page misalignment is even required to enable a successful attack, since with a very low misalignment (\eg 42), we cannot control a full \SI{4}{\kilo\byte} page. 
For instance with a misalignment of \SIx{42} bytes we only control \SIx{4071} ($8125-4096+42$) bytes of the page to leak the record data (leak page).
Therefore, the misalignment needs to be large enough to control a full leak page. 
Furthermore, the misalignment is unknown and we need to leak the misalignment of the page.
This can be done via a remote memory-deduplication attack by trying different offsets until the correct one is found.
The misalignment is the start of the record $r_{AT}$ (\cf~\cref{fig:innodb_page}).
We determine the minimal necessary misalignment $\textit{offset}_{r_{AT}}$ after the initial reorganization:

{\footnotesize
\begin{myalign*}
\textrm{offset}_{r_{AT}} mod 4096 + |r_{AT}| - 1 &\geq 4096 * 2 \\
\textrm{offset}_{r_{AT}} mod 4096 &\geq 4096 * 2 + 1 - |r_{AT}|  \\
\textrm{offset}_{r_{AT}} &\geq 68.
\end{myalign*}
}%

Hence, we need a misalignment of $r_{AT}$ of at least \SI{68}{\byte}.
Furthermore, in case the record header moves to the leak page, we would have another unknown value to leak. 
Therefore, the misalignment needs to be smaller or equal to $4096-|r_{AT_{header}}|$.
Each record is preceded by a record header ($r_{AT_{header}}$), which is maximum \SIx{27} bytes in our scenario.
We have the probability of \SI{0.66}{\percent} that the $r_{AT_{header}}$ moves to the leak page and \SI{1.66}{\percent} that the misalignment is less than \SIx{68}.
We derive the maximum leakage rate for a InnoDB index page as:
{\footnotesize%
\begin{myalign*}
  max\_leakage\_possible = \\ \min(|r_{AT}| - 1 - |r_{AT_{header}}|  - \\ \textrm{offset}_{r_{T}} \, mod \, 4096 + |r_{AT}| - 2 \cdot 4096 + 1), |r_{T}|, 4096)
\end{myalign*}
}%

Hence, we can leak up to a full size of $r_{T} \leq 4096$ if we leak the misalignment and the requirements for $\textrm{offset}_{r_{AT}}$ hold.
This limits the leakage potential of the InnoDB attack.

\section*{MariaDB User Table.}\label{sec:appendix:mariadb}
User table used in attack on InnoDB used in MariaDB~\Cref{fig:user_table}.

\begin{figure}[h]
    \centering
    \resizebox{0.35\hsize}{!}{
      \tikzsetnextfilename{mysql-table}
      \begin{tikzpicture}[]

\draw[fill=red!10] (4,5) rectangle +(4,-1.25) node [pos=.5] {\parbox{3.5cm}{\centering \textbf{Id:int}}};
\draw[fill=blue!10] (4,3.75) rectangle +(4,-1.25) node [pos=.5] {\parbox{3.5cm}{\centering username:varchar(200)}};
\draw[fill=blue!10] (4,2.5) rectangle +(4,-1.25) node [pos=.5] {\parbox{3.5cm}{\centering password::varchar(200)}};
\draw[fill=blue!10] (4,1.25) rectangle +(4,-1.25) node [pos=.5] {\parbox{3.5cm}{\centering image::mediumblob}};

\node at (6,6) {\textbf{User Table}};

\end{tikzpicture}
    }
    \caption{MariaDB user table, which is susceptible to a remote memory-deduplication attack.}
    \label{fig:user_table}
\end{figure}

\section*{Memcached Eviction.}\label{sec:appendix:memcached-eviction}
The attacks on Memcached rely on the assumption that an attacker can reliably trigger copy-on-write pagefaults by updating the same item.
However, one problem that can occur in the Memcached attack is that another user gets the free deduplicated page assigned instead of the attacker.
Therefore, it is a race between the attacker and potential other users to get the page and then trigger the copy-on-write page fault on the deduplicated page.
Another issue is whether the pages stays cached in Memcached for a longer period until the deduplication by the operating system happens, \ie \SIx{15} minutes on Windows.
The eviction totally depends on the size of the Memcached instance itself.

In this experiment, we validate how long an entry is cached in an Memcached instance with different memory limits.
First, we launch a new memcached instance and load as many entries into the instance until the memory limit is exhausted and the instance has to evict existing entries.

\begin{figure}[t]
    \centering
    \resizebox{\hsize}{!}{
        \tikzsetnextfilename{memcached-eviction}
        \begin{tikzpicture}
\begin{axis}[
style={font=\footnotesize},
xlabel={Memory Limit [MB]},
ylabel={Execution Time [s]},
y label style={align=center,text width=1.5cm},
width=0.9\hsize,
xmin=1,
height=3cm,
legend style={at={(0.0,1.0)}, anchor=north west, legend columns=2, font=\tiny,draw=none,fill=none},
]

\addplot+[thick,mark=o,blue] table[x=Limit,y=Time,col sep=comma] {data/memcached-eviction.csv};

\end{axis}
\end{tikzpicture}
    }
    \caption{Average execution time in seconds ($n=10$) until a newly added entry is evicted from memcached depending on its given memory limit.}
    \label{fig:memcached_eviction}
\end{figure}
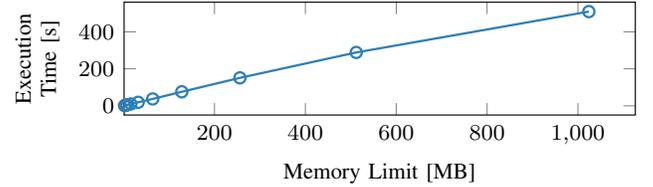

Then, we add a new entry and probe how many seconds this entry remains cached while we simultaneously apply a realistic workload on the instance.
We utilize memtier\_benchmark~\cite{Redis2013MemtierBenchmark}, spawning 4 threads that simultaneously write and read entries from the memcached instance using a gaussian access pattern with an average of \SIx{671121} ops/sec and an average bandwith of \SI{268.26}{\mega\byte/\second}.
\Cref{fig:memcached_eviction} illustrates after how many seconds on average the added entry is evicted from the memcached instance ($n=10$).
\Cref{fig:memcached_eviction} illustrates the eviction time for different Memcached node sizes.
As can be seen, the larger the node is, the longer it takes to evict a certain item.

\end{document}